\pdfoutput=1
\documentclass[11pt,twoside,a4paper,pdftex,cmspaper,final,collab]{cms-tdr}

\begin{document}\cmsNoteHeader{CFT-09-015}
%%%%%%%%%%%%%%%%%%%%%%%%%%%%%%%%%%%%%%%%%%%%%%%%%%%%%%%%%%%%%%%%%%%%
%
%  Common definitions
%
%  N.B. use of \providecommand rather than \newcommand means
%       that a definition is ignored if already specified
%
%                                              L. Taylor 18 Feb 2005
%%%%%%%%%%%%%%%%%%%%%%%%%%%%%%%%%%%%%%%%%%%%%%%%%%%%%%%%%%%%%%%%%%%%

%%%%%%%%%%%%%%%%%%%%%%%%%%%%%%%%%%%%%%%%%%%%%%%%%%%%%%%%%%%%%%%%%%%%
%
% Hyphenations (only need to add here if you get a nasty word break)
%
\hyphenation{env-iron-men-tal}%    just an example
\hyphenation{had-ron-i-za-tion}
\hyphenation{cal-or-i-me-ter}
\hyphenation{de-vices}
%
% Hyphenations-end
%
%
\RCS$Revision: 1.110 $
\RCS$Date: 2010/01/04 00:33:35 $
\RCS$Name:  $
%%%%%%%%%%%%% ptdr definitions %%%%%%%%%%%%%%%%%%%%%
%%%%%%%%%%%%%%%%%%%%%%%%%%%%%%%%%%%%%%%%%%%%%%%%%%%%%%%%%%%%%%%%%%%%
%
%  Common definitions
%
%  N.B. use of \providecommand rather than \newcommand means
%       that a definition is ignored if already specified
%
%                                              L. Taylor 18 Feb 2005
%%%%%%%%%%%%%%%%%%%%%%%%%%%%%%%%%%%%%%%%%%%%%%%%%%%%%%%%%%%%%%%%%%%%

% Some shorthand
% turn off italics
\newcommand {\etal}{\mbox{et al.}\xspace} %et al. - no preceding comma
\newcommand {\ie}{\mbox{i.e.}\xspace}     %i.e.
\newcommand {\eg}{\mbox{e.g.}\xspace}     %e.g.
\newcommand {\etc}{\mbox{etc.}\xspace}     %etc.
\newcommand {\vs}{\mbox{\sl vs.}\xspace}      %vs.
\newcommand {\mdash}{\ensuremath{\mathrm{-}}} % for use within formulas

% some terms whose definition we may change
\newcommand {\Lone}{Level-1\xspace} % Level-1 or L1 ?
\newcommand {\Ltwo}{Level-2\xspace}
\newcommand {\Lthree}{Level-3\xspace}

% Some software programs (alphabetized)
\providecommand{\ACERMC} {\textsc{AcerMC}\xspace}
\providecommand{\ALPGEN} {{\textsc{alpgen}}\xspace}
\providecommand{\CHARYBDIS} {{\textsc{charybdis}}\xspace}
\providecommand{\CMKIN} {\textsc{cmkin}\xspace}
\providecommand{\CMSIM} {{\textsc{cmsim}}\xspace}
\providecommand{\CMSSW} {{\textsc{cmssw}}\xspace}
\providecommand{\COBRA} {{\textsc{cobra}}\xspace}
\providecommand{\COCOA} {{\textsc{cocoa}}\xspace}
\providecommand{\COMPHEP} {\textsc{CompHEP}\xspace}
\providecommand{\EVTGEN} {{\textsc{evtgen}}\xspace}
\providecommand{\FAMOS} {{\textsc{famos}}\xspace}
\providecommand{\GARCON} {\textsc{garcon}\xspace}
\providecommand{\GARFIELD} {{\textsc{garfield}}\xspace}
\providecommand{\GEANE} {{\textsc{geane}}\xspace}
\providecommand{\GEANTfour} {{\textsc{geant4}}\xspace}
\providecommand{\GEANTthree} {{\textsc{geant3}}\xspace}
\providecommand{\GEANT} {{\textsc{geant}}\xspace}
\providecommand{\HDECAY} {\textsc{hdecay}\xspace}
\providecommand{\HERWIG} {{\textsc{herwig}}\xspace}
\providecommand{\HIGLU} {{\textsc{higlu}}\xspace}
\providecommand{\HIJING} {{\textsc{hijing}}\xspace}
\providecommand{\IGUANA} {\textsc{iguana}\xspace}
\providecommand{\ISAJET} {{\textsc{isajet}}\xspace}
\providecommand{\ISAPYTHIA} {{\textsc{isapythia}}\xspace}
\providecommand{\ISASUGRA} {{\textsc{isasugra}}\xspace}
\providecommand{\ISASUSY} {{\textsc{isasusy}}\xspace}
\providecommand{\ISAWIG} {{\textsc{isawig}}\xspace}
\providecommand{\MADGRAPH} {\textsc{MadGraph}\xspace}
\providecommand{\MCATNLO} {\textsc{mc@nlo}\xspace}
\providecommand{\MCFM} {\textsc{mcfm}\xspace}
\providecommand{\MILLEPEDE} {{\textsc{millepede}}\xspace}
\providecommand{\ORCA} {{\textsc{orca}}\xspace}
\providecommand{\OSCAR} {{\textsc{oscar}}\xspace}
\providecommand{\PHOTOS} {\textsc{photos}\xspace}
\providecommand{\PROSPINO} {\textsc{prospino}\xspace}
\providecommand{\PYTHIA} {{\textsc{pythia}}\xspace}
\providecommand{\SHERPA} {{\textsc{sherpa}}\xspace}
\providecommand{\TAUOLA} {\textsc{tauola}\xspace}
\providecommand{\TOPREX} {\textsc{TopReX}\xspace}
\providecommand{\XDAQ} {{\textsc{xdaq}}\xspace}

%  Experiments
\newcommand {\DZERO}{D\O\xspace}     %etc.

% Measurements and units...

\newcommand{\de}{\ensuremath{^\circ}}
\newcommand{\ten}[1]{\ensuremath{\times \text{10}^\text{#1}}}
\newcommand{\unit}[1]{\ensuremath{\text{\,#1}}\xspace}
\newcommand{\mum}{\ensuremath{\,\mu\text{m}}\xspace}
\newcommand{\micron}{\ensuremath{\,\mu\text{m}}\xspace}
\newcommand{\cm}{\ensuremath{\,\text{cm}}\xspace}
\newcommand{\mm}{\ensuremath{\,\text{mm}}\xspace}
\newcommand{\mus}{\ensuremath{\,\mu\text{s}}\xspace}
\newcommand{\keV}{\ensuremath{\,\text{ke\hspace{-.08em}V}}\xspace}
\newcommand{\MeV}{\ensuremath{\,\text{Me\hspace{-.08em}V}}\xspace}
\newcommand{\GeV}{\ensuremath{\,\text{Ge\hspace{-.08em}V}}\xspace}
\newcommand{\TeV}{\ensuremath{\,\text{Te\hspace{-.08em}V}}\xspace}
\newcommand{\PeV}{\ensuremath{\,\text{Pe\hspace{-.08em}V}}\xspace}
\newcommand{\keVc}{\ensuremath{{\,\text{ke\hspace{-.08em}V\hspace{-0.16em}/\hspace{-0.08em}c}}}\xspace}
\newcommand{\MeVc}{\ensuremath{{\,\text{Me\hspace{-.08em}V\hspace{-0.16em}/\hspace{-0.08em}c}}}\xspace}
\newcommand{\GeVc}{\ensuremath{{\,\text{Ge\hspace{-.08em}V\hspace{-0.16em}/\hspace{-0.08em}c}}}\xspace}
\newcommand{\TeVc}{\ensuremath{{\,\text{Te\hspace{-.08em}V\hspace{-0.16em}/\hspace{-0.08em}c}}}\xspace}
\newcommand{\keVcc}{\ensuremath{{\,\text{ke\hspace{-.08em}V\hspace{-0.16em}/\hspace{-0.08em}c}^\text{2}}}\xspace}
\newcommand{\MeVcc}{\ensuremath{{\,\text{Me\hspace{-.08em}V\hspace{-0.16em}/\hspace{-0.08em}c}^\text{2}}}\xspace}
\newcommand{\GeVcc}{\ensuremath{{\,\text{Ge\hspace{-.08em}V\hspace{-0.16em}/\hspace{-0.08em}c}^\text{2}}}\xspace}
\newcommand{\TeVcc}{\ensuremath{{\,\text{Te\hspace{-.08em}V\hspace{-0.16em}/\hspace{-0.08em}c}^\text{2}}}\xspace}

\newcommand{\pbinv} {\mbox{\ensuremath{\,\text{pb}^\text{$-$1}}}\xspace}
\newcommand{\fbinv} {\mbox{\ensuremath{\,\text{fb}^\text{$-$1}}}\xspace}
\newcommand{\nbinv} {\mbox{\ensuremath{\,\text{nb}^\text{$-$1}}}\xspace}
\newcommand{\percms}{\ensuremath{\,\text{cm}^\text{$-$2}\,\text{s}^\text{$-$1}}\xspace}
\newcommand{\lumi}{\ensuremath{\mathcal{L}}\xspace}
\newcommand{\Lumi}{\ensuremath{\mathcal{L}}\xspace}%both upper and lower
%
% Need a convention here:
\newcommand{\LvLow}  {\ensuremath{\mathcal{L}=\text{10}^\text{32}\,\text{cm}^\text{$-$2}\,\text{s}^\text{$-$1}}\xspace}
\newcommand{\LLow}   {\ensuremath{\mathcal{L}=\text{10}^\text{33}\,\text{cm}^\text{$-$2}\,\text{s}^\text{$-$1}}\xspace}
\newcommand{\lowlumi}{\ensuremath{\mathcal{L}=\text{2}\times \text{10}^\text{33}\,\text{cm}^\text{$-$2}\,\text{s}^\text{$-$1}}\xspace}
\newcommand{\LMed}   {\ensuremath{\mathcal{L}=\text{2}\times \text{10}^\text{33}\,\text{cm}^\text{$-$2}\,\text{s}^\text{$-$1}}\xspace}
\newcommand{\LHigh}  {\ensuremath{\mathcal{L}=\text{10}^\text{34}\,\text{cm}^\text{$-$2}\,\text{s}^\text{$-$1}}\xspace}
\newcommand{\hilumi} {\ensuremath{\mathcal{L}=\text{10}^\text{34}\,\text{cm}^\text{$-$2}\,\text{s}^\text{$-$1}}\xspace}

% Some usual physics terms

\newcommand{\zp}{\ensuremath{\mathrm{Z}^\prime}\xspace}

% SM (still to be classified)

\newcommand{\kt}{\ensuremath{k_{\mathrm{T}}}\xspace}
\newcommand{\BC}{\ensuremath{{B_{\mathrm{c}}}}\xspace}
\newcommand{\bbarc}{\ensuremath{{\overline{b}c}}\xspace}
\newcommand{\bbbar}{\ensuremath{{b\overline{b}}}\xspace}
\newcommand{\ccbar}{\ensuremath{{c\overline{c}}}\xspace}
\newcommand{\JPsi}{\ensuremath{{J}/\psi}\xspace}
\newcommand{\bspsiphi}{\ensuremath{B_s \to \JPsi\, \phi}\xspace}
\newcommand{\AFB}{\ensuremath{A_\mathrm{FB}}\xspace}
\newcommand{\EE}{\ensuremath{e^+e^-}\xspace}
\newcommand{\MM}{\ensuremath{\mu^+\mu^-}\xspace}
\newcommand{\TT}{\ensuremath{\tau^+\tau^-}\xspace}
\newcommand{\wangle}{\ensuremath{\sin^{2}\theta_{\mathrm{eff}}^\mathrm{lept}(M^2_\mathrm{Z})}\xspace}
\newcommand{\ttbar}{\ensuremath{{t\overline{t}}}\xspace}
\newcommand{\stat}{\ensuremath{\,\text{(stat.)}}\xspace}
\newcommand{\syst}{\ensuremath{\,\text{(syst.)}}\xspace}
% these moved to similar defs
%\newcommand{\Etmiss}{\ensuremath{E_{\mathrm{T}\!{\rm miss}}}}
%\newcommand{\VEtmiss}{\ensuremath{{\vec E}_{\mathrm{T}\!{\rm miss}}}}

%%%  E-gamma definitions
\newcommand{\HGG}{\ensuremath{\mathrm{H}\to\gamma\gamma}}
\newcommand{\gev}{\GeV}
\newcommand{\GAMJET}{\ensuremath{\gamma + \mathrm{jet}}}
\newcommand{\PPTOJETS}{\ensuremath{\mathrm{pp}\to\mathrm{jets}}}
\newcommand{\PPTOGG}{\ensuremath{\mathrm{pp}\to\gamma\gamma}}
\newcommand{\PPTOGAMJET}{\ensuremath{\mathrm{pp}\to\gamma +
\mathrm{jet}
}}
\newcommand{\MH}{\ensuremath{\mathrm{M_{\mathrm{H}}}}}
\newcommand{\RNINE}{\ensuremath{\mathrm{R}_\mathrm{9}}}
\newcommand{\DR}{\ensuremath{\Delta\mathrm{R}}}

% Physics symbols ...

\newcommand{\PT}{\ensuremath{p_{\mathrm{T}}}\xspace}
\newcommand{\pt}{\ensuremath{p_{\mathrm{T}}}\xspace}
\newcommand{\ET}{\ensuremath{E_{\mathrm{T}}}\xspace}
\newcommand{\HT}{\ensuremath{H_{\mathrm{T}}}\xspace}
\newcommand{\et}{\ensuremath{E_{\mathrm{T}}}\xspace}
\newcommand{\Em}{\ensuremath{E\!\!\!/}\xspace}
\newcommand{\Pm}{\ensuremath{p\!\!\!/}\xspace}
\newcommand{\PTm}{\ensuremath{{p\!\!\!/}_{\mathrm{T}}}\xspace}
\newcommand{\ETm}{\ensuremath{E_{\mathrm{T}}^{\mathrm{miss}}}\xspace}
\newcommand{\MET}{\ensuremath{E_{\mathrm{T}}^{\mathrm{miss}}}\xspace}
\newcommand{\ETmiss}{\ensuremath{E_{\mathrm{T}}^{\mathrm{miss}}}\xspace}
\newcommand{\VEtmiss}{\ensuremath{{\vec E}_{\mathrm{T}}^{\mathrm{miss}}}\xspace}

%%%%%%
% From Albert
%

\newcommand{\ga}{\ensuremath{\gtrsim}}
\newcommand{\la}{\ensuremath{\lesssim}}
\newcommand{\swsq}{\ensuremath{\sin^2\theta_W}\xspace}
\newcommand{\cwsq}{\ensuremath{\cos^2\theta_W}\xspace}
\newcommand{\tanb}{\ensuremath{\tan\beta}\xspace}
\newcommand{\tanbsq}{\ensuremath{\tan^{2}\beta}\xspace}
\newcommand{\sidb}{\ensuremath{\sin 2\beta}\xspace}
\newcommand{\alpS}{\ensuremath{\alpha_S}\xspace}
\newcommand{\alpt}{\ensuremath{\tilde{\alpha}}\xspace}

\newcommand{\QL}{\ensuremath{Q_L}\xspace}
\newcommand{\sQ}{\ensuremath{\tilde{Q}}\xspace}
\newcommand{\sQL}{\ensuremath{\tilde{Q}_L}\xspace}
\newcommand{\ULC}{\ensuremath{U_L^C}\xspace}
\newcommand{\sUC}{\ensuremath{\tilde{U}^C}\xspace}
\newcommand{\sULC}{\ensuremath{\tilde{U}_L^C}\xspace}
\newcommand{\DLC}{\ensuremath{D_L^C}\xspace}
\newcommand{\sDC}{\ensuremath{\tilde{D}^C}\xspace}
\newcommand{\sDLC}{\ensuremath{\tilde{D}_L^C}\xspace}
\newcommand{\LL}{\ensuremath{L_L}\xspace}
\newcommand{\sL}{\ensuremath{\tilde{L}}\xspace}
\newcommand{\sLL}{\ensuremath{\tilde{L}_L}\xspace}
\newcommand{\ELC}{\ensuremath{E_L^C}\xspace}
\newcommand{\sEC}{\ensuremath{\tilde{E}^C}\xspace}
\newcommand{\sELC}{\ensuremath{\tilde{E}_L^C}\xspace}
\newcommand{\sEL}{\ensuremath{\tilde{E}_L}\xspace}
\newcommand{\sER}{\ensuremath{\tilde{E}_R}\xspace}
\newcommand{\sFer}{\ensuremath{\tilde{f}}\xspace}
\newcommand{\sQua}{\ensuremath{\tilde{q}}\xspace}
\newcommand{\sUp}{\ensuremath{\tilde{u}}\xspace}
\newcommand{\suL}{\ensuremath{\tilde{u}_L}\xspace}
\newcommand{\suR}{\ensuremath{\tilde{u}_R}\xspace}
\newcommand{\sDw}{\ensuremath{\tilde{d}}\xspace}
\newcommand{\sdL}{\ensuremath{\tilde{d}_L}\xspace}
\newcommand{\sdR}{\ensuremath{\tilde{d}_R}\xspace}
\newcommand{\sTop}{\ensuremath{\tilde{t}}\xspace}
\newcommand{\stL}{\ensuremath{\tilde{t}_L}\xspace}
\newcommand{\stR}{\ensuremath{\tilde{t}_R}\xspace}
\newcommand{\stone}{\ensuremath{\tilde{t}_1}\xspace}
\newcommand{\sttwo}{\ensuremath{\tilde{t}_2}\xspace}
\newcommand{\sBot}{\ensuremath{\tilde{b}}\xspace}
\newcommand{\sbL}{\ensuremath{\tilde{b}_L}\xspace}
\newcommand{\sbR}{\ensuremath{\tilde{b}_R}\xspace}
\newcommand{\sbone}{\ensuremath{\tilde{b}_1}\xspace}
\newcommand{\sbtwo}{\ensuremath{\tilde{b}_2}\xspace}
\newcommand{\sLep}{\ensuremath{\tilde{l}}\xspace}
\newcommand{\sLepC}{\ensuremath{\tilde{l}^C}\xspace}
\newcommand{\sEl}{\ensuremath{\tilde{e}}\xspace}
\newcommand{\sElC}{\ensuremath{\tilde{e}^C}\xspace}
\newcommand{\seL}{\ensuremath{\tilde{e}_L}\xspace}
\newcommand{\seR}{\ensuremath{\tilde{e}_R}\xspace}
\newcommand{\snL}{\ensuremath{\tilde{\nu}_L}\xspace}
\newcommand{\sMu}{\ensuremath{\tilde{\mu}}\xspace}
\newcommand{\sNu}{\ensuremath{\tilde{\nu}}\xspace}
\newcommand{\sTau}{\ensuremath{\tilde{\tau}}\xspace}
\newcommand{\Glu}{\ensuremath{g}\xspace}
\newcommand{\sGlu}{\ensuremath{\tilde{g}}\xspace}
\newcommand{\Wpm}{\ensuremath{W^{\pm}}\xspace}
\newcommand{\sWpm}{\ensuremath{\tilde{W}^{\pm}}\xspace}
\newcommand{\Wz}{\ensuremath{W^{0}}\xspace}
\newcommand{\sWz}{\ensuremath{\tilde{W}^{0}}\xspace}
\newcommand{\sWino}{\ensuremath{\tilde{W}}\xspace}
\newcommand{\Bz}{\ensuremath{B^{0}}\xspace}
\newcommand{\sBz}{\ensuremath{\tilde{B}^{0}}\xspace}
\newcommand{\sBino}{\ensuremath{\tilde{B}}\xspace}
\newcommand{\Zz}{\ensuremath{Z^{0}}\xspace}
\newcommand{\sZino}{\ensuremath{\tilde{Z}^{0}}\xspace}
\newcommand{\sGam}{\ensuremath{\tilde{\gamma}}\xspace}
\newcommand{\chiz}{\ensuremath{\tilde{\chi}^{0}}\xspace}
\newcommand{\chip}{\ensuremath{\tilde{\chi}^{+}}\xspace}
\newcommand{\chim}{\ensuremath{\tilde{\chi}^{-}}\xspace}
\newcommand{\chipm}{\ensuremath{\tilde{\chi}^{\pm}}\xspace}
\newcommand{\Hone}{\ensuremath{H_{d}}\xspace}
\newcommand{\sHone}{\ensuremath{\tilde{H}_{d}}\xspace}
\newcommand{\Htwo}{\ensuremath{H_{u}}\xspace}
\newcommand{\sHtwo}{\ensuremath{\tilde{H}_{u}}\xspace}
\newcommand{\sHig}{\ensuremath{\tilde{H}}\xspace}
\newcommand{\sHa}{\ensuremath{\tilde{H}_{a}}\xspace}
\newcommand{\sHb}{\ensuremath{\tilde{H}_{b}}\xspace}
\newcommand{\sHpm}{\ensuremath{\tilde{H}^{\pm}}\xspace}
\newcommand{\hz}{\ensuremath{h^{0}}\xspace}
\newcommand{\Hz}{\ensuremath{H^{0}}\xspace}
\newcommand{\Az}{\ensuremath{A^{0}}\xspace}
\newcommand{\Hpm}{\ensuremath{H^{\pm}}\xspace}
\newcommand{\sGra}{\ensuremath{\tilde{G}}\xspace}
\newcommand{\mtil}{\ensuremath{\tilde{m}}\xspace}
\newcommand{\rpv}{\ensuremath{\rlap{\kern.2em/}R}\xspace}
\newcommand{\LLE}{\ensuremath{LL\bar{E}}\xspace}
\newcommand{\LQD}{\ensuremath{LQ\bar{D}}\xspace}
\newcommand{\UDD}{\ensuremath{\overline{UDD}}\xspace}
\newcommand{\Lam}{\ensuremath{\lambda}\xspace}
\newcommand{\Lamp}{\ensuremath{\lambda'}\xspace}
\newcommand{\Lampp}{\ensuremath{\lambda''}\xspace}
\newcommand{\spinbd}[2]{\ensuremath{\bar{#1}_{\dot{#2}}}\xspace}

\newcommand{\MD}{\ensuremath{{M_\mathrm{D}}}\xspace}% ED mass
\newcommand{\Mpl}{\ensuremath{{M_\mathrm{Pl}}}\xspace}% Planck mass
\newcommand{\Rinv} {\ensuremath{{R}^{-1}}\xspace}

%%%%%%%%%%%%%%%%%%%%%%%%%%%%%%%%%%%%%%%%%%%%%%%%%%%%%%%%%%%%%%%%%%%%
%
% Hyphenations (only need to add here if you get a nasty word break)
%
\hyphenation{en-viron-men-tal}%    just an example

%%%%%%%%%%%%%%%  Title page %%%%%%%%%%%%%%%%%%%%%%%%
\cmsNoteHeader{09-015}
\title{Precise Mapping of the Magnetic Field in the  CMS Barrel
  Yoke using Cosmic Rays}

\author[]{The CMS Collaboration}

\date{\today}

\abstract{
  The CMS detector is designed around a large 4~T superconducting
  solenoid, enclosed in a 12\ 000-tonne steel return yoke.
  A detailed map of the magnetic field is required for the
  accurate simulation and reconstruction of physics events in the CMS
  detector, not only in the inner tracking region inside the solenoid
  but also in the large and  complex structure of the steel yoke, which
  is instrumented with muon chambers.
  Using a large sample of cosmic muon events collected by CMS in 2008, the
  field in the steel of the barrel yoke has been determined with a
  precision of 3 to 8\% depending on the location.
}

\hypersetup{%
pdfauthor={The CMS Collaboration},%
pdftitle={Precise Mapping of the Magnetic Field in the CMS Barrel Yoke using Cosmic Rays},%
pdfsubject={CMS},%
pdfkeywords={CMS, physics, magnetic field}}

\maketitle

%%%%%%%%%%%%%%%%%%%%%%%%%%%%%%%%  Begin text %%%%%%%%%%%%%%%%%%%%%%%%%%%%%

\section{Introduction}
\begin{figure}[hbt]
  \begin{center}
    \includegraphics[width=0.45\textwidth]{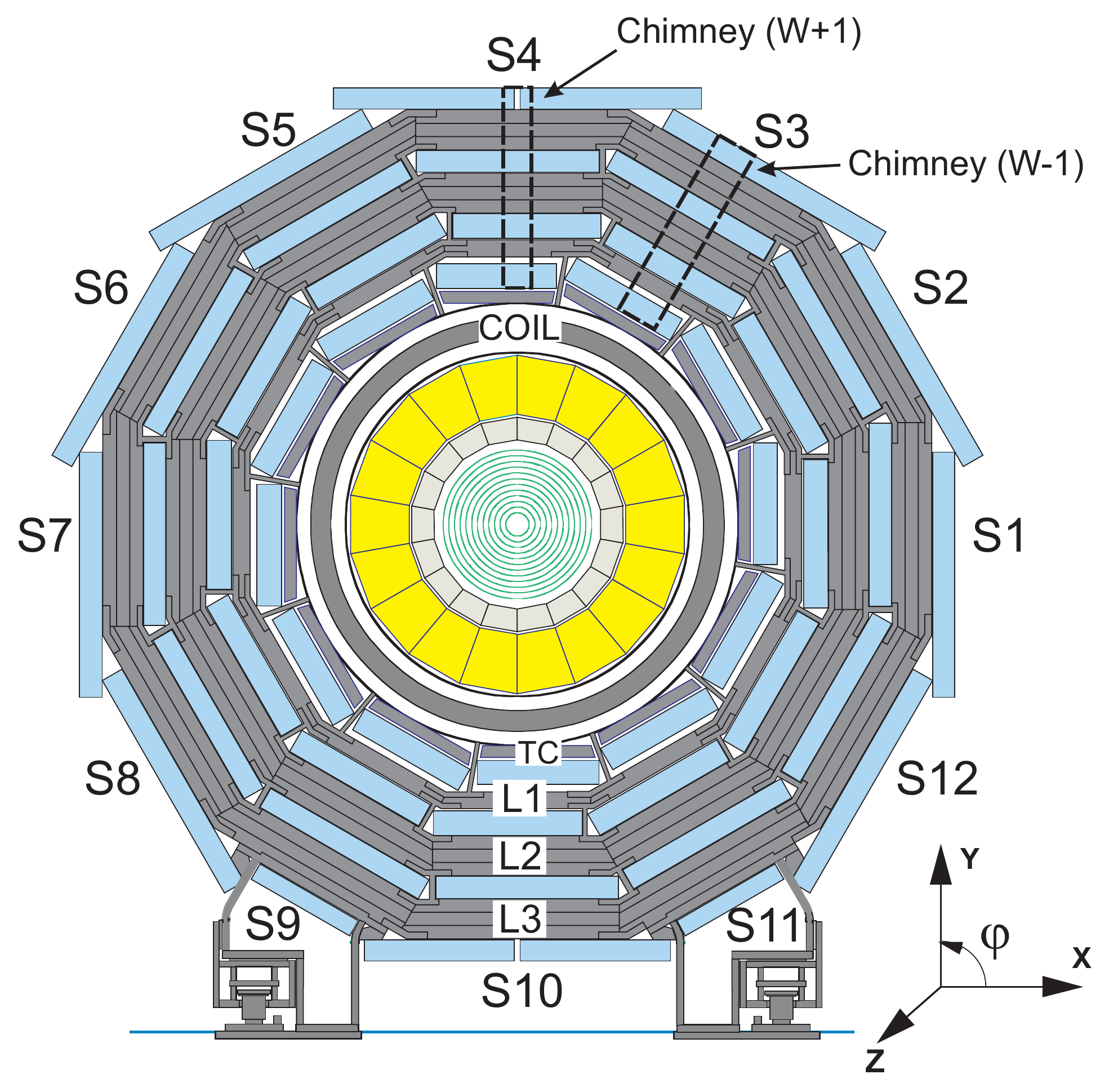}%
    \hfill
    \includegraphics[width=0.54\textwidth]{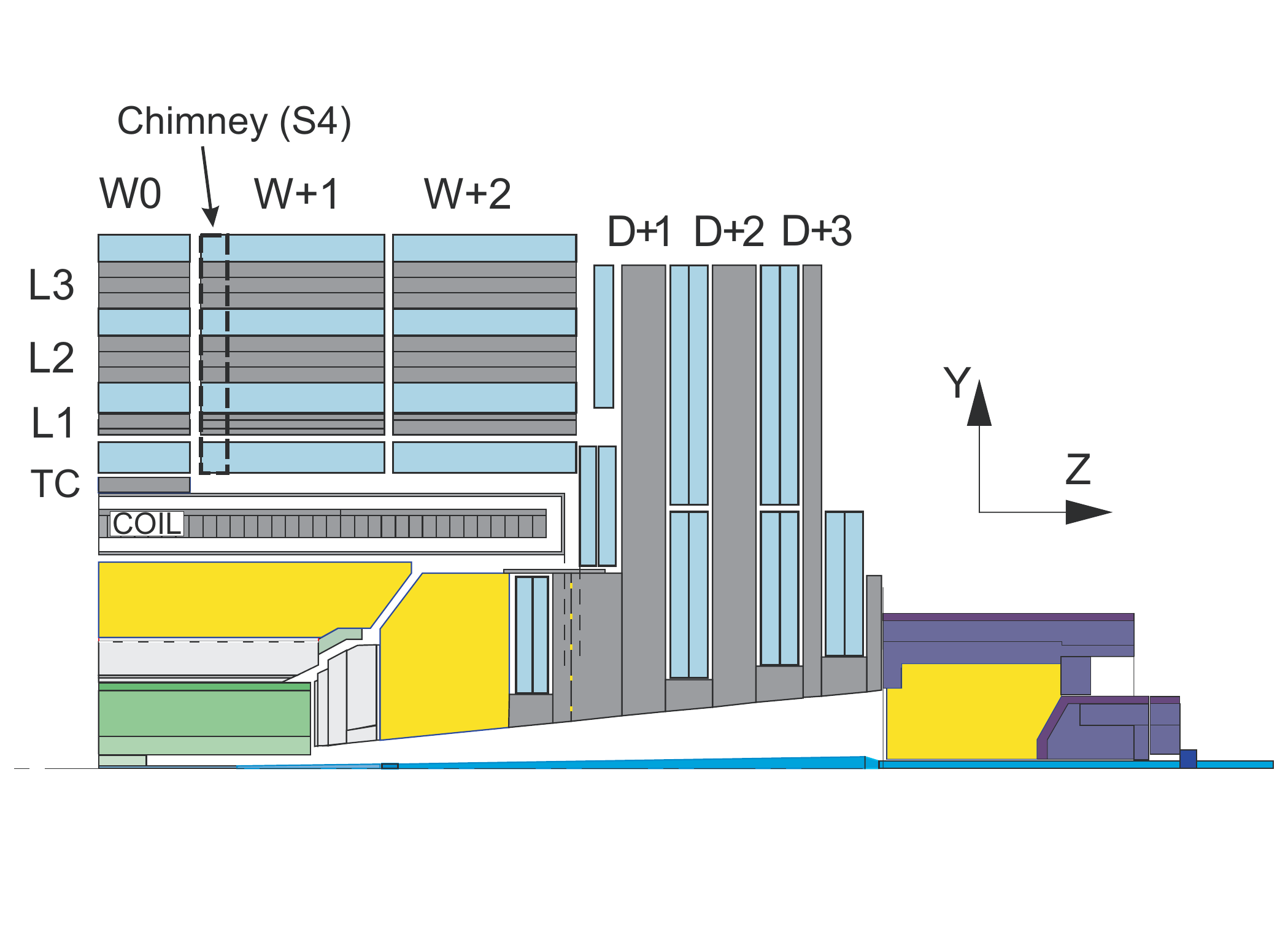}%
    \hfill
    \caption{Schematic views of the CMS detector, with the numbering
      convention for azimuthal sectors (S), wheels (W), barrel yoke layers (L)
      and endcap disks (D). ``TC'' is the ``tail catcher'', an
      additional steel layer present in the central barrel wheel only.
      Left: transverse view at $z=0$. Right:
      longitudinal view of one quarter of the detector.
      Wheels on the $z<0$ side (not shown) are labeled W-1 and
      W-2. The location of the ``chimneys'' described in
      Section~\ref{sec:TOSCA} is indicated with boxes.}
    \label{fig:Sectors}
  \end{center}
\end{figure}

The Compact Muon Solenoid (CMS)~\cite{CMS} is a general-purpose
detector whose main goal is to explore physics at the TeV scale by
exploiting the proton-proton collisions provided by the Large
Hadron Collider (LHC)~\cite{LHC} at CERN. Its distinctive features include
a 4~T superconducting solenoid with a free bore of a diameter of 6~m
and a length of 12.5~m, enclosed inside a \mbox{12\ 000}-tonne yoke made of
common structural steel~\cite{MagnetTDR}. The geometry of CMS is shown in
Fig.~\ref{fig:Sectors}. The yoke is composed of
five three-layered dodecagonal barrel wheels
and three endcap disks at each end. 
In the barrel region the innermost yoke layer is 295~mm
thick and each of the two outermost ones is 630~mm thick.
The yoke contributes to only 8\% of the central magnetic flux
density; its main role is to increase the field homogeneity in the
tracker volume and to reduce the stray field by returning the magnetic
flux of the
solenoid. In addition, the steel plates play the role of absorber for the four
interleaved layers (``stations'') of muon chambers, which 
provide for a measurement of the muon
momentum independent of the inner tracking system. 

The CMS collaboration has decided to operate the magnet at a
central magnetic flux density of 3.8~T. After the
first years of operation, once the aging of the coil is better understood, the
collaboration may decide to operate the magnet at 4~T. 
%All necessary tests and field maps have been also made at this value. 

In order to allow accurate reconstruction of track parameters and
Monte Carlo simulation of events, a detailed map of the magnetic
field in the entire volume of the CMS detector is needed.
The accuracy of the magnetic field map is
discussed in Section~\ref{sec:requirements}.
Sections~\ref{sec:TOSCA} and~\ref{sec:implementation} describe the
computation of the CMS field map and its implementation in the CMS
software framework, respectively.

Several techniques have been adopted to directly measure and to monitor
the magnetic flux density $B$
in the various parts of the CMS detector, as discussed
in Section~\ref{sec:probes}.
In particular, the field in the central volume of the solenoid
was mapped with very good precision.
However, measurements with probes outside the solenoid
are not sufficient to constrain the field map inside the steel of the
yoke to the level of precision required.

During October-November 2008, the CMS Collaboration conducted a month-long
data-taking exercise known as the Cosmic Run At Four Tesla (CRAFT),
with the goal of commissioning the experiment for extended operation
\cite{CRAFTGeneral}.
With all installed detector systems participating, CMS recorded 270 million
cosmic ray triggered events with the solenoid at a central magnetic
flux density of 3.8~T.
Using these data it was possible for the first time to 
probe the magnetic field in the steel of the return yoke using
reconstructed muon tracks. The field in the different parts of
the barrel yoke was measured and correction factors for the field map
were obtained, as
described in Section~\ref{sec:CRAFT}.

This paper does not cover the study of the field in the endcap
yoke. That
analysis is more challenging, since
the number of cosmic muons traversing both the inner tracker and
the endcap stations is limited for geometrical reasons.

%%%%%%%%%%%%%%%%%%%%%%%%%%%%%%%%%%%%%%%%%%%%%%%%%%%%%%%%%%%%%%%%%%%%%%
\section{Accuracy of the Magnetic Field Map}
\label{sec:requirements}

The CMS silicon tracker, the central detector for charged particle track reconstruction,
is located centrally inside the superconducting coil of
the magnet of the CMS detector.
Within that region, the field has a high strength and is
relatively homogeneous.
As discussed in Section~\ref{sec:probes}, the field in the tracker
volume has been mapped with an accuracy better than 0.1\%.
This precision is crucial for physics analyses
as it allows accurate measurements of charged particle track
parameters near the interaction vertex.

\begin{figure}
  \begin{center}
    \includegraphics[width=0.49\textwidth]{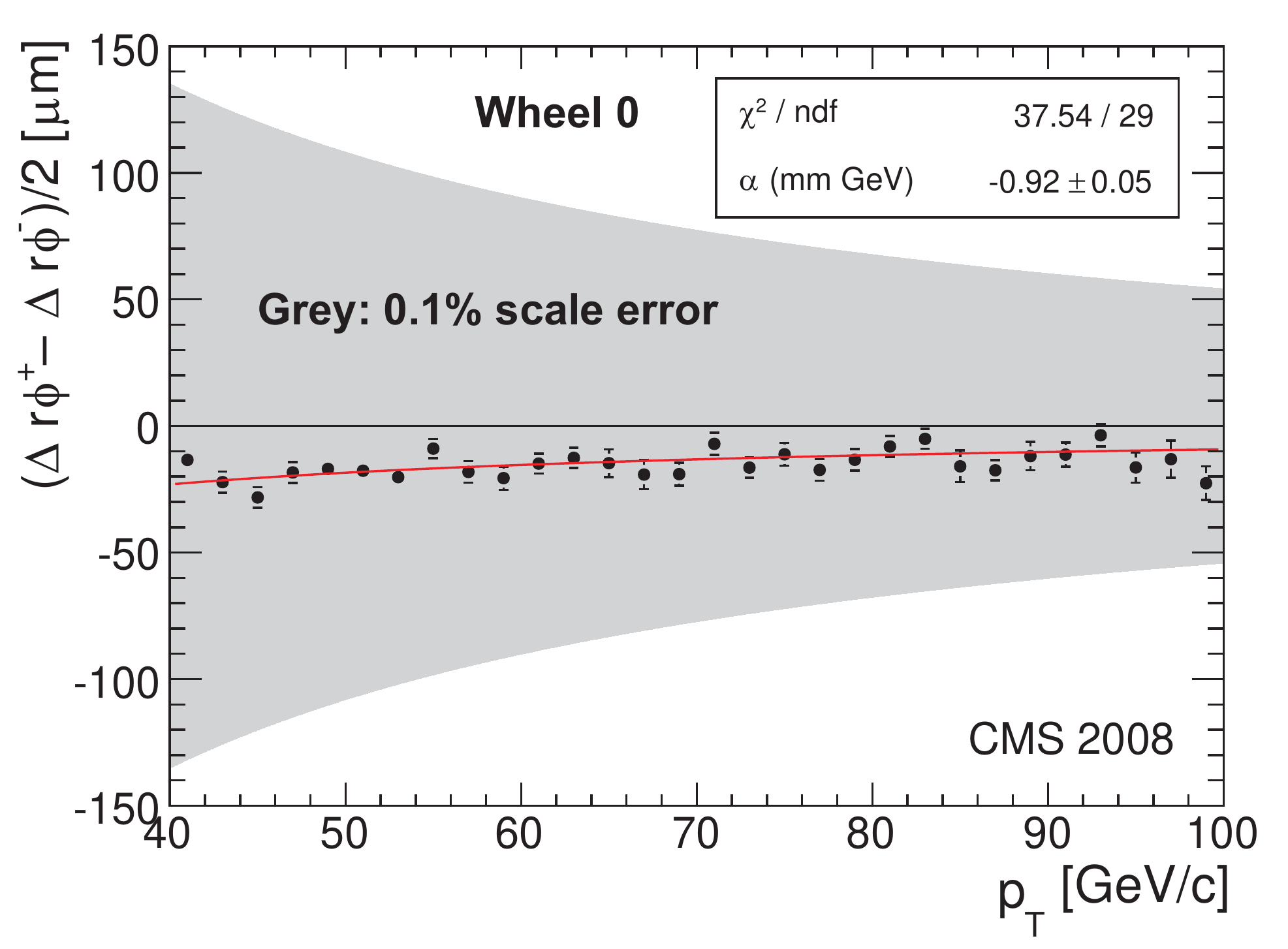}
    \includegraphics[width=0.49\textwidth]{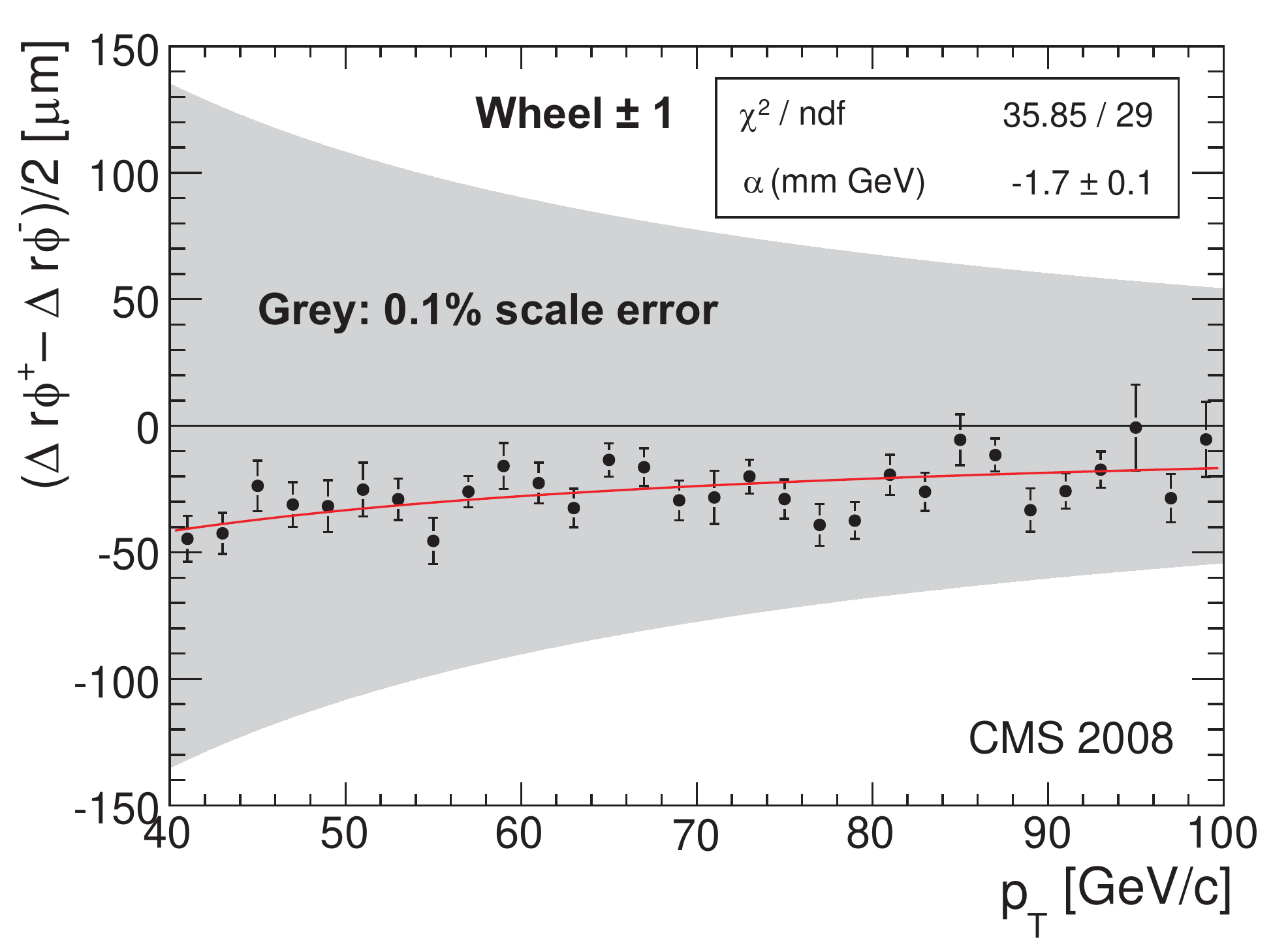}
    \caption{Residual distances, in the bending plane, between the
      extrapolation of the tracker tracks and the measurements in the
      first muon station, for the CRAFT cosmic muon data set, as a
      function of the transverse momentum.
      Left: wheel 0. Right: wheels $\pm1$.
      The shaded area shows the expected effect of a
      0.1\% distortion of the field map in the region between the
      inner tracker and the coil.  The solid line represents a fit to
      the function $\alpha/p_T$.
      }
    \label{fig:accuracyTkMB1}
  \end{center}
\end{figure}

Outside the tracker volume, the field map is calculated with the
finite-element computation described in Section~\ref{sec:TOSCA}.
Between the tracker and the first barrel muon station the magnetic field
integral is dominated by the relatively homogeneous field inside
the coil ($R<3$~m). In this region, the accuracy of the computation is
confirmed by its agreement within 0.06\% with fixed Nuclear Magnetic
Resonance (NMR) probes located at $R=2.91$~m (cf. Section~\ref{sec:mapping}).
An additional test of the accuracy of the field map inside
the solenoid is given by the comparison of the predicted and observed
bending of cosmic ray tracks between the silicon tracker and the first
muon station. For this purpose, cosmic muons reconstructed with the
inner tracker are extrapolated to the first muon station using the
calculated field map. The residual between the measured and the
extrapolated positions, $\Delta$, is computed separately for positive and negative
muons, and their charge-antisymmetric combination
($(\Delta^{\mu+}-\Delta^{\mu-})/2$) is used to suppress the
possible effect of a residual misalignment. The result
is plotted in Fig.~\ref{fig:accuracyTkMB1}.
The expected effect of a 0.1\% distortion of the field
map in the region located between the tracker and the first muon
station, relative to the field inside the tracker,
is shown as a reference. We conclude that the map
describes the field in this region within this accuracy.

The muon momentum and charge are mostly determined by the curvature of
the reconstructed track in the tracker~\cite{PTDR1}.
Only at very high transverse momentum a track fit combining hits in
the tracker and in the muon system is expected to improve the resolution,
due to the long additional lever arm with high magnetic field between
the outer layer of the tracker and the muon system.
Studies with
cosmic muons show that the most critical factor in this global fit is
alignment~\cite{CFTMuonReco, CFTMuTrackAlignment}.

A measurement of the muon momentum can also be obtained exclusively using
information from the muon chambers.
This ``stand-alone'' muon reconstruction is used in
the hardware-based \mbox{Level-1} muon trigger
and in the first stage of the High-Level Trigger~\cite{HLT}.
The resolution that can be obtained in this case is limited by
multiple scattering, by the finite resolution of the muon chambers, and
by their alignment.
Ideally, the magnetic field map should be known with sufficient
accuracy so that a possible biasing effect on the momentum scale is
small compared to the resolution of the momentum measurement.  While a
systematic momentum bias is an uncertainty of a fundamentally
different nature than the detector resolution, this requirement
ensures that the efficiency of matching tracker tracks to
muon hits for identification and reconstruction purposes is not
reduced, and that the sharpness of trigger turn-on curves is not
affected by spatially inhomogeneous inaccuracies of the field map.

\begin{figure}
  \begin{center}
    \includegraphics[width=0.44\textwidth]{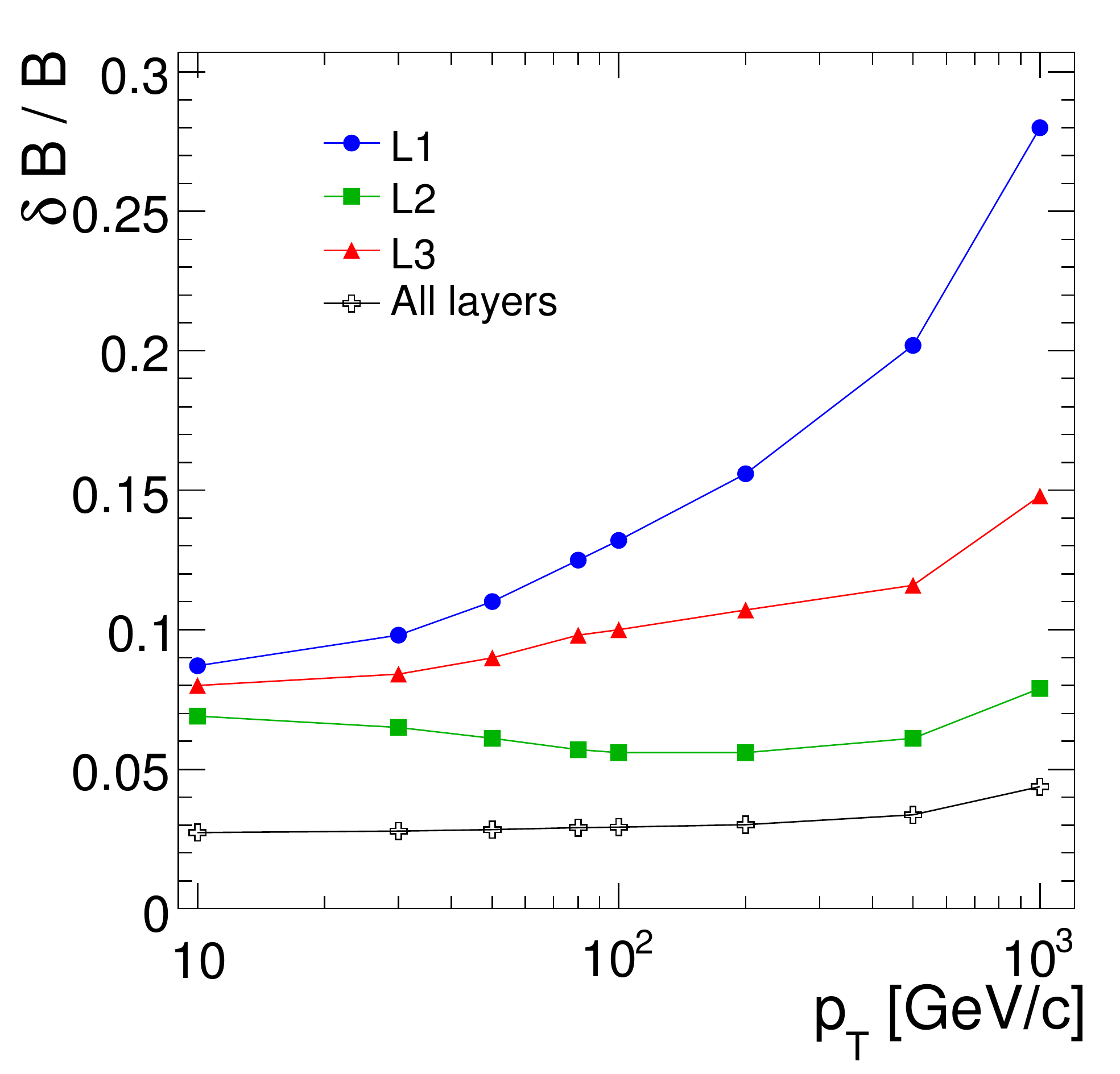}
    \caption{
      The systematic uncertainty in the $B$ scale that corresponds to a systematic
      uncertainty in the determination of the muon momentum equal to
      one tenth of the momentum resolution, for muons reconstructed using
      the barrel muon spectrometer alone and without a vertex
      constraint.
      The different curves correspond to a coherent scaling of $B$ in a
      single yoke layer, without changing the field in other layers
      (curves labeled ``L1'', ``L2'', ``L3'') and to a scaling
      in all layers with the same factor (curve labeled ``All layers'').
      }
    \label{fig:accuracy}
  \end{center}
\end{figure}

An analytical calculation was used to estimate how the fitted muon
momentum is affected by the inaccuracies of the magnetic field
map in the yoke~\cite{ThesisBianchini}.
The largest effects occur if only the bending power provided by
the return field in the yoke is considered, without constraining the fitted
muon tracks to the beamspot (``vertex constraint'').
Figure~\ref{fig:accuracy}
shows at which point the relative systematic bias on the momentum
due to limited knowledge of the field is ten times smaller than
the resolution of the stand-alone momentum measurement deriving from
hit resolution and multiple scattering, under the assumption of
perfect detector alignment.
The most stringent condition on $B$ in a single  layer is found
in the second layer of the return yoke, where the magnetic field
systematic uncertainty is required to be below 5\%
at intermediate momentum. If all layers are affected by a fully correlated
scaling factor, this value becomes 3\%.
Thus, to ensure that the systematic uncertainty due to the inaccuracy
of the field map is negligible, the benchmark is set at 3\% for the
overall scale uncertainty and at 5\% for the scale in
individual plates in the barrel return yoke.

These limits are conservative, as they are
obtained for the extreme case of a fit with no vertex constraint. The
constrained fit profits from the lever arm and bending power between
the vertex and
the first  layer of the muon system, improving the precision of the fit
for muons originating from the interaction region and reducing the
reliance on the accuracy of the field in the return yoke by one order
of magnitude.
%reliance on the field in the return yoke, so that the requirements on
%the accuracy of the field can be relaxed by one order of magnitude.

%%%%%%%%%%%%%%%%%%%%%%%%%%%%%%%%%%%%%%%%%%%%%%%%%%%%%%%%%%%%%%%%%%%%%%
\section{Finite-Element Model of the CMS Magnetic Field}
\label{sec:TOSCA}
In order to establish a map of the CMS magnetic field,
the CMS solenoid and yoke were modeled using the
TOSCA finite element program~\cite{TOSCA}. We summarize here the main
features of this model, which is  described in detail elsewhere~\cite{CERN:042}.

The orientation of the CMS reference frame and the naming
conventions used in this paper for the different parts of the steel
yoke are shown in Fig.~\ref{fig:Sectors}.
CMS uses a right-handed coordinate system, with the origin at the nominal
collision point, the $x$-axis pointing to the centre of the LHC, the $y$-axis
pointing up (perpendicular to the LHC plane), and the $z$-axis along the
anticlockwise-beam direction.
The azimuthal angle, $\phi$, is measured from the positive $x$-axis
in the $x$-$y$ plane.

The steel yoke is composed of 12 azimuthal sectors, and is therefore to
a good approximation 12-fold $\phi$-symmetric,
except for a few features:
\begin{itemize}
\item the presence of radial passages in steel slabs (``chimneys'')
  to route cryogenic and electric
  connections in sector 3 of wheel $-1$ (with a depth along $z$ of
  39~cm and 84~cm wide)
  and sector 4 of wheel +1 (39~cm deep and 54~cm wide);
%plates have Dz=2.536m and Drphi=2.551m (L1),  3.033m (L2), 3.582m (L3)
\item the presence of supporting feet in sectors 9 and 11;
\item the presence of the carts supporting the endcap disks;
\item the presence of a steel plate on the floor under the detector.
\end{itemize}

The entire length of CMS along the $z$-axis had to be modeled since the
winding of the CMS solenoid is not exactly $z$-symmetric, affecting the field
in the inner tracker region. However, due to limitations on the
maximum number of nodes in the TOSCA mesh only the $x>0$ half of the
CMS detector was modeled. This choice allows an approximated map
of the field in the entire detector to be obtained using the 12-fold symmetry of the yoke
with a special treatment of the features described above, as will be
discussed in the next section.

Two different yoke configurations have been modeled. The first
(``surface model'') represents the
status of the CMS detector during the mapping campaign of 2006, that
was performed in the surface hall. At that time, the small outermost
endcap steel disk shown in Fig.~\ref{fig:Sectors}
was not present in the negative-$z$ endcap,
and neither were any ferromagnetic parts beyond
$|z|>10.86$~m, i.e., the forward hadron calorimeters and
the shielding of the LHC magnets. This configuration is used only for
comparison with the 2006 measurements.
The second (``underground model'') represents the final setup of
CMS in the experimental hall.

\begin{figure}
  \begin{center}
    \includegraphics[width=0.5\textwidth,trim=115 0 170 0, clip]{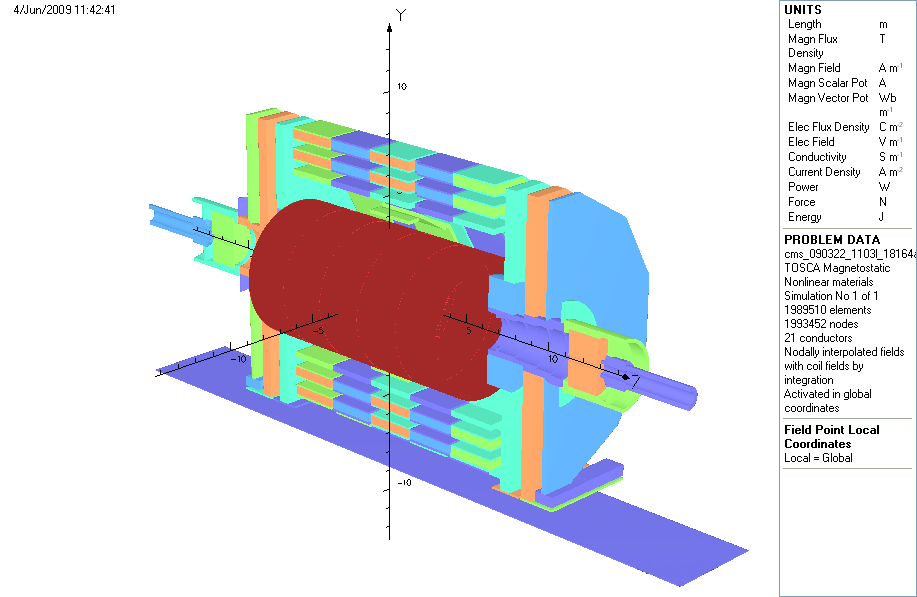} %fig/ToscaVolumes
    \caption{Representation of the magnetic elements included
        in the TOSCA model (underground configuration).}
    %TODO: provide: NA
    \label{fig:Tosca}
  \end{center}
\end{figure}

\begin{figure}
  \begin{center}
    \includegraphics[width=\textwidth,trim=0 0 193 0, clip]{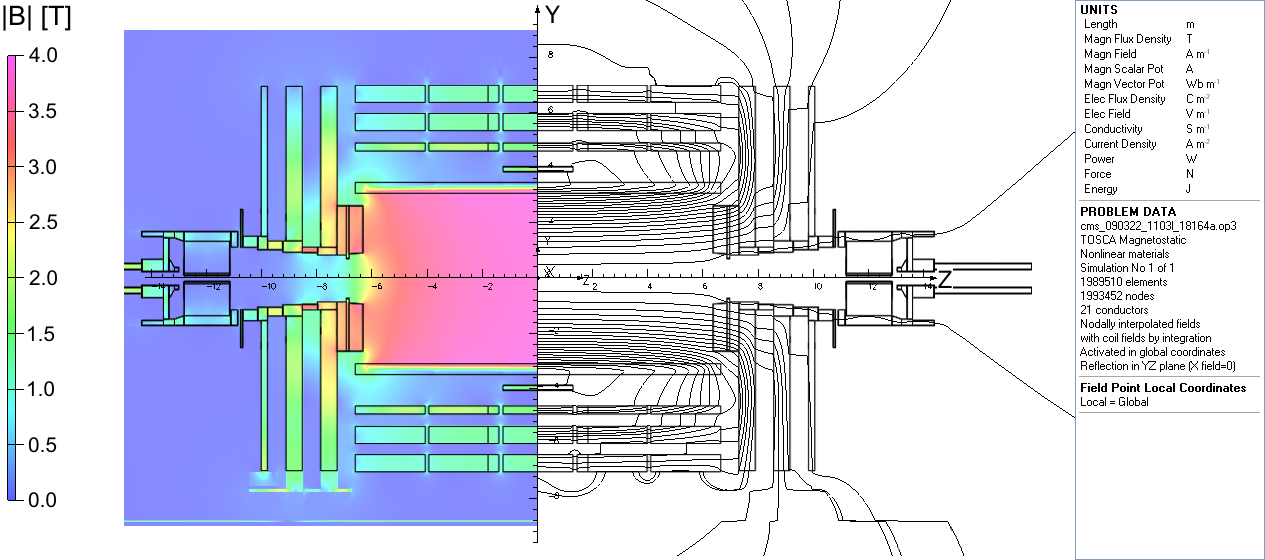}

    \caption{Value of $|B|$ (left) and field lines (right) predicted on
      a longitudinal section of the CMS detector, for the underground
      model at a central magnetic flux density of 3.8~T. Each field
      line represents a magnetic flux increment of 6~Wb.}
    \label{fig:FieldLines}
  \end{center}
\end{figure}

\begin{figure}
  \begin{center}
    \includegraphics[width=0.75\textwidth]{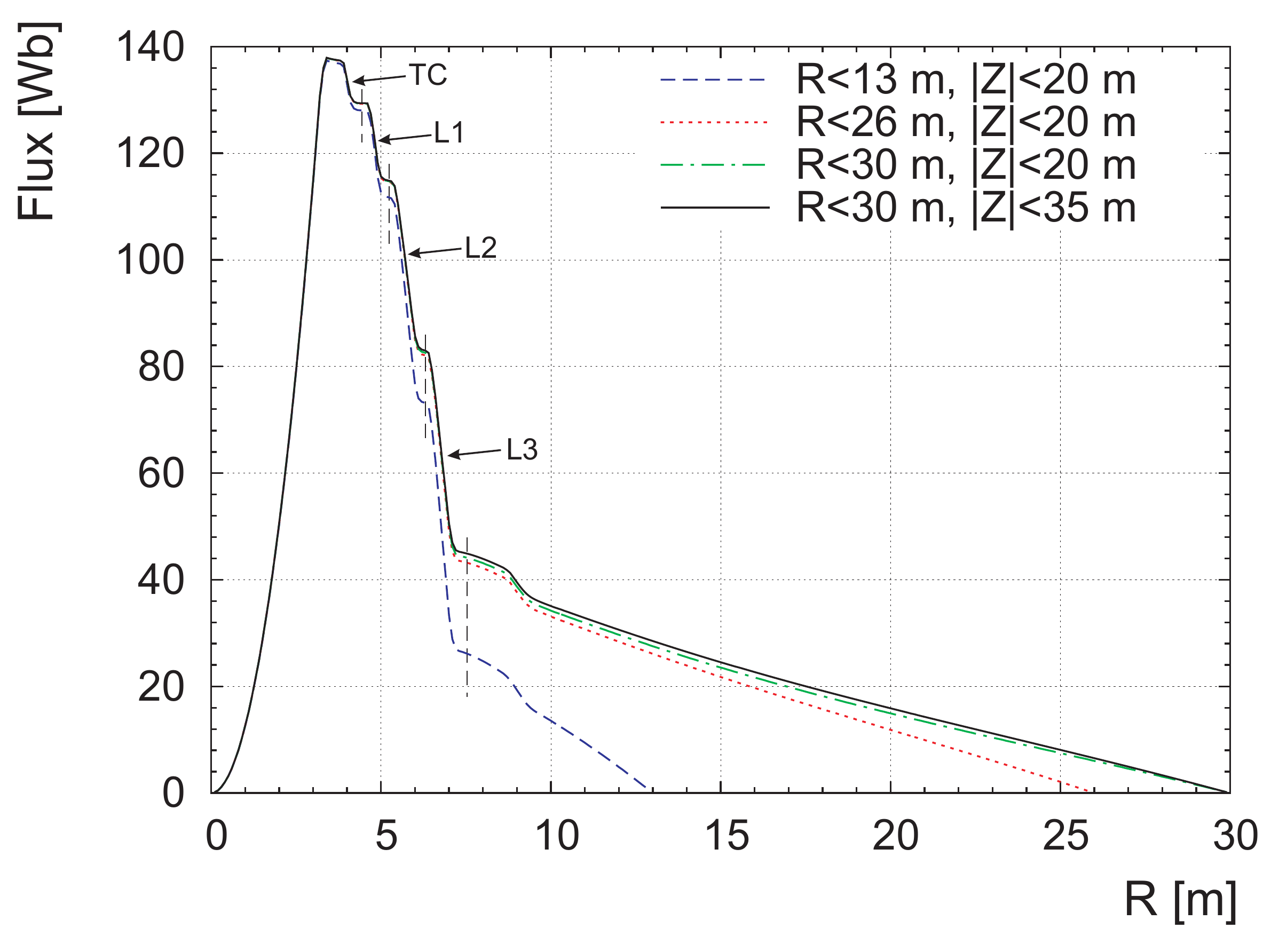}
    \caption{Magnetic flux through a disk of
        radius $R$ in the transverse plane at $z = 0$~m,
        as predicted by the TOSCA
        model for the surface configuration and 4~T operating
        conditions, computed within different enclosing regions. The
        vertical dashed lines indicate the location of the four barrel
        muon stations.}
    \label{fig:flux}
  \end{center}
\end{figure}

Figure~\ref{fig:Tosca} shows a representation of the model geometry.
The predicted magnetic flux density  on a longitudinal section of the CMS
detector is shown in Fig.~\ref{fig:FieldLines}.
 Approximately two
thirds of the magnetic flux return through the barrel yoke, half of which
enters directly into the barrel without passing through the endcap disks.
One third of the total flux escapes radially, returning outside the
steel yoke.
For this reason,
particular care has to be taken in modeling boundary positions. Truly open
boundaries cannot be specified with TOSCA. The simplest way to
approximate open boundaries is to compute the field in a large
region enclosing the solenoid and yoke. The effect of different
choices for the enclosing volume on the same TOSCA model is shown in
Fig.~\ref{fig:flux}. A small enclosing
region (e.g., $R < 13$~m) forces too much flux to return in the yoke,
causing a distortion in the region instrumented with muon detectors
($4 < R < 7.4$~m).
By increasing the enclosing region to $R<26$~m, the total flux
returning through the
yoke is reduced by about 15\%. A further increase to a region $R<30$~m
gives an additional reduction of only about 1\%.
%             R<13   R<26    R<30    R<30,Z<35
%max          137.4  137.86  137.89 137.91
%MB3(R=6.31)  73.06  82.06   82.55  82.96
%MB4(R=7.53): 26.16  43.2    44.10  44.91
%L3           46.9   38.86   38.45  38.05
%                -17%     -1.1%  -1.0%
%max-MB4      111.2  94.66   93.79  93
%                -14.9%   -0.92% -0.84%
%Air R(from 7m) 6    19      23
%                9.0x  1.47x
The region enclosing the model cannot be enlarged indefinitely without
reducing the precision of the calculation, due to the
limitations in the number of mesh points in TOSCA.
Except where otherwise specified, the models used in the
rest of this paper are computed in the largest of the regions considered in
Fig.~\ref{fig:flux}. Any residual effect due to boundary positions,
as well as the effect of additional magnetic material in the
experimental cavern (in particular, reinforcement steel in the concrete
walls, that is not included in the model) has to be estimated and
calibrated out with real data, as discussed in Section~\ref{sec:CRAFT}.

%%%%%%%%%%%%%%%%%%%%%%%%%%%%%%%%%%%%%%%%%%%%%%%%%%%%%%%%%%%%%%%%%%%%%%
\section{The Field Map for Simulation and Reconstruction}
\label{sec:implementation}
Simulation and reconstruction of events in the CMS detector require
knowledge of the magnetic field in the entire detector, both in the
inner tracking region and in the complex configuration of the steel
return yoke. Besides accuracy, computing efficiency of the map
interface is a key requirement, as the map is accessed intensively
during the on-line reconstruction in the High-Level Trigger.

An optimized interface~\cite{CHEP04} to the data obtained from
the TOSCA computation described in the previous section
%in section~\ref{sec:TOSCA}
was developed in the framework of the standard CMS software, to be used
for Monte Carlo simulation, High-Level Trigger, and offline reconstruction.

The map is implemented using a dedicated, volume-based geometry model.
The detector is subdivided in volumes, which correspond to the volumes
used in TOSCA to define the magnetic properties of the different
materials. Their boundaries correspond to the field
discontinuities due to changes in the magnetic permeability of
different materials. 
The field within each volume is
therefore continuous, allowing track extrapolation algorithms to
optimize the step size. Volumes are organized in a hierarchical
structure optimized for fast global search. In addition, caching and
navigation techniques allow simulation and track extrapolation
algorithms to minimize the number of global volume searches.

Within each volume, the field is obtained either by interpolation of
the values computed with TOSCA on a three-dimen\-sional grid adapted to the
volume shape so that it is regular in an appropriate coordinate system,
or by parameterizations of the TOSCA computation when available.
Although the different volumes constituting the CMS detector can use
specific grid tables, symmetries can be exploited to reduce the memory
footprint of the map.
In particular, a compact field map of the entire CMS
detector has been produced exploiting the 12-fold $\phi$-symmetry of
the yoke, with special treatment of the main $\phi$-asymmetric features
described in Section~\ref{sec:TOSCA}.
For this purpose, specific grid tables are used for the three barrel
steel layers in the sectors containing chimneys (S3, S4).
Specific tables are also used in the bottom sectors (S9, S10, S11) of
the outermost barrel steel layer and of the endcap disks,  that are
influenced by the presence of the feet and carts.
The field in all other sectors is obtained with appropriate
rotations from the grid tables of sector 1, which is chosen as it is
well separated from the $\phi$-asymmetric features of the yoke.
This choice represents a compromise between accuracy, memory footprint,
and complexity of the map.

%%%%%%%%%%%%%%%%%%%%%%%%%%%%%%%%%%%%%%%%%%%%%%%%%%%%%%%%%%%%%%%%%%%%%%
\section{Direct Field Measurements}
\label{sec:probes}

Detailed measurements of the magnetic field using various sensors were
performed in 2006 in the surface hall.
These results
have been complemented with new measurements in the cavern during
the CRAFT campaign. In this section, the measured values are compared
with the latest TOSCA computations.

\subsection{Measurements within the tracker volume}
\label{sec:mapping}
The field inside the coil was measured in the surface hall, before
lowering CMS into the experimental cavern and before the installation
of the tracker and of the electromagnetic calorimeter. A pneumo-mechanical field mapper was used to scan a
cylinder of radius 1.724~m and length 7~m, providing precise
measurements for \mbox{33\ 840} space points. A
detailed description of the device and of the results is given
in Ref.~\cite{CERN:040}.

The field has been measured at several values of $B_0$, the central
magnetic flux density: 2, 3, 3.5, 3.8 and 4~T.
Comparison with the TOSCA model
shows excellent agreement with the measurements at
$B_0=4$~T, with a discrepancy smaller than 5~mT.
The model reproduces a small $z-$asymmetry of
the magnetic flux density 
caused by one missing turn of the coil out of the 2180 designed turns,
which was discovered during the field mapping campaign.
%Numbers are taken Fig. 3 and 4 of~\cite{CERN:040}; these are for the
%coil axis and for the largest measured radius.

For optimal representation of these data, a $\phi$-symmetric parametrization
has been implemented using an expansion of
the magnetic scalar potential over spherical harmonics represented
in a cylindrical coordinate system, satisfying the Laplace
equation~\cite{thesisVassili}.
The expansion coefficients fitted from the measured field values
resulted in a map of the field in the volume comprising the inner tracker
that agrees with the $\phi$-averaged measurements within 0.2 mT.
This fit includes a refinement of the estimated gains and offsets of
the Hall probes, with the overall scale set by NMR probe measurements,
confirming and further improving the initial $5\times10^{-4}$ accuracy
of the calibration.
This analysis confirms the excellent quality of the field mapper measurements.
The parametrization is available for processing as an option for
applications that require the highest possible accuracy and for
studies of systematic uncertainties related to the magnetic field.
The default setting for simulation, tracking, and High-Level Trigger
is to use the standard map based on TOSCA (cf. Section~\ref{sec:TOSCA}),
which is sufficiently precise and computationally less demanding.

Simulation studies have shown that the different magnetic environment inside
the experimental cavern has effects on the overall scale of the
field in the yoke. However, the effect on the field inside the solenoid is
negligible. This has been confirmed by comparing measurements with
fixed NMR probes installed near the inner wall of the superconducting
coil cryostat, taken at the same coil current on surface and
underground (see Table~\ref{tab:NMR}). At the time of the underground
measurements, the inner tracker and the electromagnetic calorimeter
were installed inside the solenoid.
It can be concluded that the measurements of the
field on the surface are applicable also for operation in the cavern.

\begin{table}
\caption{Comparison of the magnetic flux density measured by fixed NMR
  probes inside the solenoid on the surface and underground, for
  different coil currents. 
  Each measurement is performed with one single probe, although
  different probes are used in different operation ranges. 
  Probes are located at $R=2.91$~m, $z = -0.01$~m.
%  Two probes, with different range of operation, are used: 
%  probe A at $\phi=44.9^{\circ}$ and probe B
%  at $\phi=-135.1^{\circ}$, 
  The relative precision of the measurements is better than
  $5\times10^{-5}$.
% Estimated looking RMS of repeated measurements and given that earth
% field is ~0.00006 T.
  The prediction of the underground TOSCA model at 18\ 160~A
  ($B_0=3.8$~T) for the location of the probe at $\phi=44.9^{\circ}$, where
  the relative variation of $|B|$ is expected to be smaller 
  than $3\times10^{-4}$ for a displacement of 1~cm, is 3.9181~T.
%   The discrepancy of the TOSCA underground model at 18\ 160~A
%   with the corresponding  measured value is 2.5~mT.
% CF B_vs_I_at_R2914_upd.xls with all relevant info.
% Actual positions of probes (only A,B used here)
% A#7: X= +2.06345 m, Y= +2.0587 m, Z= -0.01 m; (Phi ~+45 deg)
% B#5: X= -2.06345 m, Y= -2.0587 m, Z= -0.01 m; (Phi ~-135 deg)
% E#7: X= -2.06345 m, Y= +2.0587 m, Z= +0.01 m; (Phi ~+135 deg)
% F#6: X= +2.06345 m, Y= -2.0587 m, Z= +0.01 m; (Phi ~-45 deg)
}
\label{tab:NMR}

\begin{center}
\begin{tabular}{c|c|cc|c}
\hline
Current [A] & Probe $\phi$ & Surface [T] & Underground [T] & $\Delta$ [T] \\
\hline
\ \ 7\ 000 & $-135.1^{\circ}$ & 1.5218 & 1.5224 & $-0.0006$ \\
\ \ 9\ 500 & $-135.1^{\circ}$ & 2.0616 & 2.0628 & $-0.0012$ \\
18\ 160 & $44.9^{\circ}$  & 3.9176 & 3.9206 & $-0.0030$ \\
\hline
\end{tabular}

\end{center}
\end{table}

In conclusion, the field in the CMS inner tracker region is known to
better than 0.1\%, and the agreement of the TOSCA model with the
measurements at $B_0=4$~T is better than 5 mT everywhere in the
mapped region.
Moreover, hit position residuals for tracks extrapolated from the tracker
to the first muon station (see Fig. 2) validate
the predicted field integral outside the mapped region.
This allows the field in the inner tracker region to be used as a
reference to probe the field in the yoke with cosmic tracks, as will
be discussed in Section~\ref{sec:CRAFT}.

\subsection{Measurements of the field inside the yoke}

\begin{figure}[hbtp]
  \begin{center}
    \includegraphics[width=0.32\textwidth]{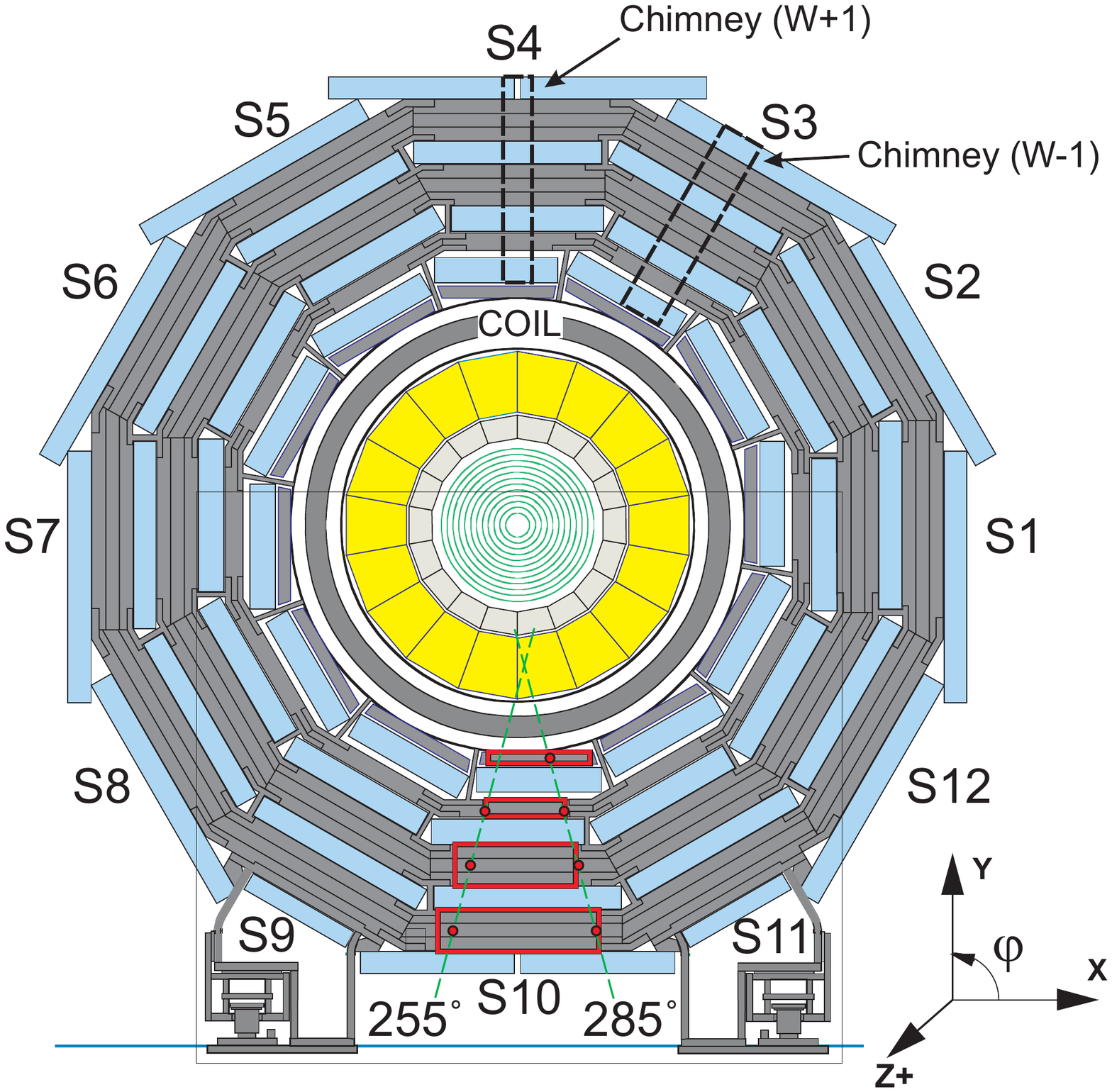}
    \hfill
    \includegraphics[width=0.64\textwidth]{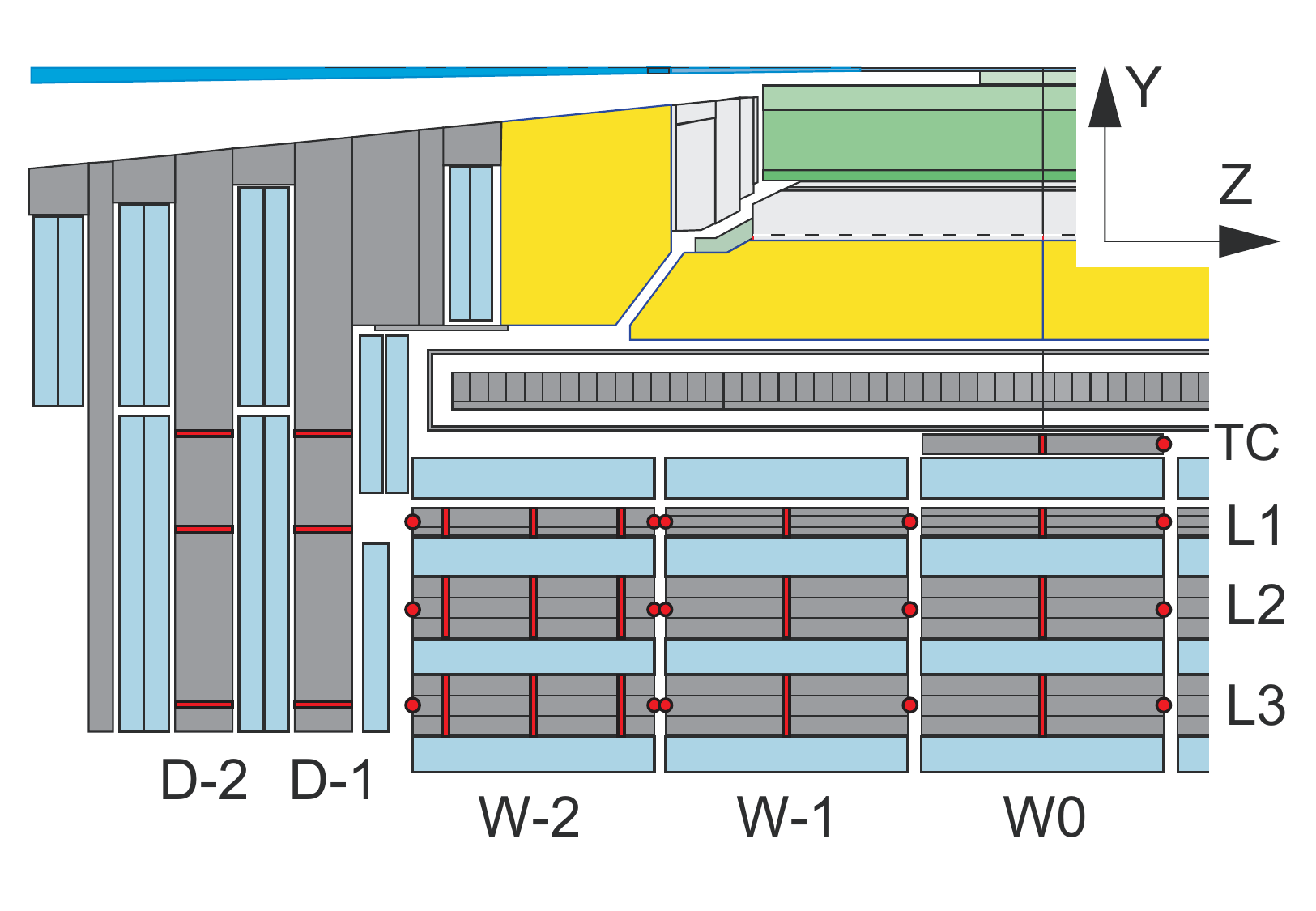}
    \hfill
    \caption{Location of flux loops (lines) and Hall probes (circles),
      projected on transverse (left) and longitudinal (right)
      sections of CMS.}
    \label{fig:HallFlux}
  \end{center}
\end{figure}

A measurement of the average magnetic flux density inside the steel
blocks of the CMS yoke was performed in the surface hall in 2006, with
a system of 22 flux loops~\cite{CERN:042} made of 315--495 turns wound
around the steel plates of sector 10 in the barrel wheels W0, W-1, and W-2,
and in the endcap disks D-1 and D-2, as shown in Fig.~\ref{fig:HallFlux}.
The areas enclosed in the flux loops vary from 0.3 to 1.7~m$^2$ in
the barrel wheels and from 0.5 to 1.2~m$^2$ in the endcap disks.
During a "fast" discharge of the coil (190~s time constant) from
the full current to zero, voltages with amplitudes up to 3--4.5~V were
induced in the loops.
The integration of the measured voltages
allows the average
magnetic flux density in the steel blocks to be derived.
At a central magnetic flux density of 4~T, the measured magnetic flux
densities in the barrel wheels vary from 0.6 to 2.1~T.
In the endcap disks, the measurements vary from 1.4 to 2.3~T.

The measured magnetic flux density is reduced by the residual
magnetization of the steel after the fast discharge.
Half of the maximum remanent field, that in the different plates
varies between 54 and 85~mT depending on the steel type, is taken as an
uncertainty on the measurement.
The precision of the voltage measurements results in an additional
uncertainty of 1.5\% on the measured flux. The total uncertainty is
therefore estimated to be 2\% to 7\% in the different loops.
A comparison with the prediction of the TOSCA surface model shows that
the average ratio of measured over calculated values is 0.97 in the barrel, with a standard
deviation of 0.07, and 0.93 in the endcap, with a standard
deviation of 0.04. For some of the measurements, large discrepancies
were found (reaching 22\% in the barrel and 14\% in the endcaps).
The current understanding of the measurements, and in particular of
the actual remanent magnetization in each yoke plate, is not
sufficient to use them to constrain the field map to a precision of
the order of 3--5\%, as required for physics analysis.
More precise measurements of the magnetic flux density within the steel
plates, also in sectors other than S10, are
possible using reconstructed tracks from cosmic rays, as discussed in
Section~\ref{sec:CRAFT}. With cosmic rays detected in the underground
experimental hall, it is also possible to observe and correct for effects due to
the magnetic environment of the cavern; this is not possible with the
available flux loop measurements, which were performed in the surface hall.

An additional measurement of the magnetic flux density in the yoke is
provided by a system of Hall probes mounted between the steel
blocks at selected places. These probes provide continuous monitoring
and are important to verify the long-term stability of the field.
The probes are mounted close to the steel blocks, but the local
magnetic flux density can only be measured in the air, and therefore
it is difficult to use these measurement to constrain the field map
within the yoke plates.
However, they can indicate overall distortions of the model, if any.

As an example, a set of measured values at a central magnetic flux density of
3.8~T is presented in Table~\ref{tab:Hall} for sensors in the barrel
wheels at $\phi=255^{\circ}$ and $285^{\circ}$.
\begin{table}
\caption{Hall probe measurements in the underground cavern at a central
  magnetic flux density $B_0=3.8$~T,
  compared with the values predicted by TOSCA for the same locations.
  Sensors are labeled according to the steel block they are located
  close to (cf. Fig.~\ref{fig:Sectors}) and their $\phi, z$ position.
%Info: numbers from run 12/11/08, except for theta285,W-2 and theta
%255,W-2,Z=-6.63 (8 rows) that are from the run of 10/10/08.
}
\label{tab:Hall}

\begin{center}
\begin{tabular}{cc|c|ccc|ccc}
\hline
\multicolumn{3}{c|}{\mbox{}} & \multicolumn{3}{c|}{$\phi=285^{\circ}$} & \multicolumn{3}{c}{$\phi=255^{\circ}$}\\
\hline
\multicolumn{2}{c|}{\mbox{}}  &  z [m] &
Data [T] & Calc [T] & Data/Calc & Data [T] & Calc [T] & Data/Calc\\
\hline
\multirow{1}{*}{TC}

& W0  &  1.273 & $-0.39$ & $-0.37$ & 1.06 & & &\\
\hline
\multirow{5}{*}{L1}
& W0  &  1.273   & $-0.96$ & $-0.97$ & 0.99 & $-0.95$ & $-0.97$ & 0.98 \\
& W-1 & $-1.418$ & $-1.02$ & $-0.97$ & 1.06 & $-0.96$ & $-0.97$ & 0.99 \\
& W-1 & $-3.964$ & $-1.13$ & $-1.12$ & 1.00 & $-1.13$ & $-1.13$ & 1.01 \\
& W-2 & $-4.079$ & $-1.14$ & $-1.13$ & 1.02 & $-1.16$ & $-1.13$ & 1.02 \\
& W-2 & $-6.625$ & $-0.26$ & $-0.31$ & 0.82 & $-0.28$ & $-0.33$ & 0.87 \\

\hline
\multirow{5}{*}{L2}
& W0  &  1.273   & $-0.89$ & $-0.95$ & 0.94 & $-0.89$ & $-0.95$ & 0.94 \\
& W-1 & $-1.418$ & $-0.90$ & $-0.96$ & 0.93 & $-0.90$ & $-0.96$ & 0.94 \\
& W-1 & $-3.964$ & $-0.87$ & $-0.90$ & 0.96 & $-0.87$ & $-0.90$ & 0.96 \\
& W-2 & $-4.079$ & $-0.87$ & $-0.90$ & 0.96 & $-0.86$ & $-0.90$ & 0.96 \\
& W-2 & $-6.625$ & $-0.34$ & $-0.41$ & 0.84 & $-0.34$ & $-0.41$ & 0.84 \\

\hline
\multirow{5}{*}{L3}
& W0  &  1.273   & $-0.82$ & $-0.88$ & 0.93 & $-0.82$ & $-0.88$ & 0.93 \\
& W-1 & $-1.418$ & $-0.82$ & $-0.89$ & 0.92 & $-0.82$ & $-0.89$ & 0.92 \\
& W-1 & $-3.964$ & $-0.75$ & $-0.78$ & 0.96 & $-0.76$ & $-0.78$ & 0.96 \\
& W-2 & $-4.079$ & $-0.75$ & $-0.78$ & 0.96 & $-0.76$ & $-0.78$ & 0.96 \\
& W-2 & $-6.625$ & $-0.37$ & $-0.49$ & 0.76 & & & \\
\hline

\end{tabular}
\end{center}
\end{table}
As shown in Fig.~\ref{fig:HallFlux}, barrel sensors are located close to
the front face of steel blocks, in the gap between two neighboring wheels.
In these gaps, the field is mostly axial and the magnetic flux density
is about two-thirds of the flux density in the center of the steel plate.

The Hall probes used in Table~\ref{tab:Hall} were calibrated at 1.4~T, with a precision better than 0.1\%.
The observed deviations can be related to the modeling of the
gaps and of local features or inhomogeneities of the steel block at the
place where the probes are mounted. To highlight a possible overall
distortion of the model, the ratio of the measured and calculated field
has been averaged for all probes located in the same gap. Results are
shown in Table~\ref{tab:HallSummary}.
In the second and third yoke layers of the barrel, the data indicate a
lower field than predicted by the TOSCA calculation.
A larger discrepancy is present in the gap between
the barrel and the endcap.

\begin{table}
\begin{center}
\caption{Ratio of the measured and calculated values
  of Table~\ref{tab:Hall}, averaged for the probes located in
  the same gap (or in $z$-symmetric gaps in case of W0/W-1). The
  largest observed difference between any two of the values that are
  averaged together is 0.08.
%  (four measurements per reported value for W0/W-1; W-1/W-2; two for
%  W-2/D-1, except for L3 where only one measurement is available)
}
\label{tab:HallSummary}
\vspace{3mm}
\begin{tabular}{l|ccc}
\hline
 & Gap W0/W$\pm$1 & Gap W-1/W-2 & Gap W-2/endcap \\
\hline
L1 & 1.01 & 1.01 & 0.84 \\
L2 & 0.94 & 0.96 & 0.84 \\
L3 & 0.92 & 0.96 & 0.76 \\
\hline
\end{tabular}
\end{center}
\end{table}

In the endcaps, Hall probes are installed between disks, to monitor
the field close to the steel surface. In this region, the field is small
(with measured values in the range between 0.01 and 0.34~T) and almost
axial, so the measured values cannot be easily related to the
field in the steel disks, which is large and radial
(cf. Fig.~\ref{fig:FieldLines}).

%%%%%%%%%%%%%%%%%%%%%%%%%%%%%%%%%%%%%%%%%%%%%%%%%%%%%%%%%%%%%%%%%%%%%%
\section{Measurements with Cosmic Ray Tracks}
\label{sec:CRAFT}
About 3\% of the muon tracks collected during CRAFT cross the acceptance of
the inner tracker.
This sample can be used to verify the accuracy of the magnetic field
map in the yoke using the information provided by the muon chambers,
and taking the precise measurement of the track momentum in the inner tracker as a reference.
This section describes a method to obtain average correction factors for the
scale of the field map in each plate of the CMS barrel yoke.

\subsection{The analysis method}
\label{sec:CRAFTBarrel}

In the barrel yoke, four stations of Drift Tube (DT) chambers are
interleaved with the three steel yoke layers, as shown in Fig.~\ref{fig:Sectors}.
Each DT chamber can measure the direction of the track in the
transverse plane ($\phi$) with a resolution of about 1.8
mrad~\cite{CFTDTPerf} based on eight measurement planes.
The track deflection in the transverse plane between two consecutive
stations, $i$ and $i+1$, is an ideal
quantity to probe the field in the yoke plates, as it is directly
related to the integral of the field along the track path:

\begin{equation}
(\phi_{i+1} -\phi_i) p_T =
-0.3\ q
\int^{i+1}_{i}\vec{u}_{\phi}\cdot\vec{B}\times d\vec{l}
\label{eq:Bintegral}
\end{equation}
where $q$ is the muon charge, $p_T$ is the muon transverse momentum in
units of GeV/$c$,
$B$ is expressed in Tesla, and $l$ in meters.

The transverse bending is dominated by the axial
component of the field, $B_z$. The azimuthal component of
the field, $B_\phi$,
is small given the cylindrical symmetry of the barrel yoke, and does
not contribute to the bending in the transverse plane.
Although the radial component, $B_r$,
contributes to Eq.~(\ref{eq:Bintegral}), its
effect is small and can be neglected, both because $B_r\ll B_z$ in the
barrel steel plates given the geometry of the yoke
(cf. Fig.~\ref{fig:FieldLines}),
and because tracks selected in this analysis have small angles with respect to the radial direction. Systematic effects due to this approximation are
discussed in Section~\ref{sec:systematics}.

In order to relate the path integral in Eq.~(\ref{eq:Bintegral}) to the
average field inside the yoke plate, the stray field in the short path
between the chambers and the steel plate can be neglected, and the
track path length in the transverse plane
can be approximated with the thickness $L$ of the steel plate. With this
approximation:

\begin{equation}
  \int^{i+1}_{i}\vec{u}_{\phi}\cdot\vec{B}\times d\vec{l} \simeq \langle B_{z}\rangle L
\label{eq:intbdl_approx}
\end{equation}
where $\langle B_{z}\rangle$ is effectively averaged along the
trajectory of the particle crossing the layer
between stations $i$ and $i+1$.

The goal of the study presented here is not to obtain
directly the value of $\langle B_{z}\rangle$ from
Eq.~(\ref{eq:intbdl_approx}), but to compare it with the same quantity
predicted
by the magnetic field map, to highlight and possibly correct for
average discrepancies.
For this purpose, track parameters reconstructed in the CMS inner
tracking system are extrapolated to the muon spectrometer, where they are
compared with the measurements of the muon chambers, for each muon separately.
The extrapolation of track parameters and of their error matrices
is performed taking into account multiple scattering and energy
loss. These were tuned to reproduce statistically the results of the
detailed GEANT4 simulation of CMS.
The simulation of the energy loss in the material between
the tracker and the muon system, amounting to about 3 GeV,
is correct to within an accuracy of 0.2~GeV, according to comparisons
with measurements~\cite{ThesisBianchini}.

The magnetic field map is used both for the measurement of the track
momentum in the inner tracker and to predict the track bending in the
extrapolation.
Given that the accuracy of the magnetic field map in the
region inside the solenoid is good, the momentum measured by the inner
tracker can be taken as reference.
A systematic difference between
the track direction measured in the DT chambers and the direction of
the extrapolated tracks can therefore be attributed to a difference
between the true magnetic field integral along the particle
path and the corresponding integral in the field map used for the
extrapolation.
Using Eq.~(\ref{eq:Bintegral}), it is possible to measure the
bias of the field integral in the steel layer placed between
two consecutive DT stations:
\begin{equation}
\Delta=[(\phi^{\rm prop}_{i+1}-\phi^{\rm data}_{i+1}) -
(\phi^{\rm prop}_{i}-\phi^{\rm data}_{i})]\cdot p_{T}%^{i+1/i}
\propto
\int^{i+1}_{i} \vec{u}_{\phi}\cdot\vec{B}^{\rm map}\times d\vec{l}
 - \int^{i+1}_{i}
 \vec{u}_{\phi}\cdot\vec{B}^{\rm true}\times d\vec{l},
\label{eq:delta}
\end{equation}
where $\phi^{\rm prop}_{i}$ and $\phi^{\rm data}_{i}$ are the bending angles at
the $i^{th}$ DT station for the propagated track and
for the track segment reconstructed in the DT chamber, respectively;
$p_{T}$ %^{i+1/i}$
is the muon
momentum, assumed constant along the path between the two stations and
obtained from the extrapolation of the inner tracker track to the
middle plane between the stations, accounting for energy loss;
and
$\int^{i+1}_{i} \vec{u}_{\phi}\cdot\vec{B}^{\rm true}\times d\vec{l}$ and
$\int^{i+1}_{i} \vec{u}_{\phi}\cdot\vec{B}^{\rm map}\times d\vec{l}$
are the true field path integral between the $i^{th}$ and the
$i+1^{th}$ DT station and the one estimated using the field map,
respectively.

The bending angles measured are potentially affected by residual
misalignment.  However, misalignment affects the measured angles of
positive and negative muons in the same direction, while a distortion of the
field map has an opposite effect on the propagated direction of tracks of
opposite charge.  The charge-antisymmetric
combination of the mean values of the distributions of $\Delta$ for
positive and negative muons crossing a given sector, in the form
$(\langle\Delta\rangle_{\mu^+}-\langle\Delta\rangle_{\mu^-})/2$,
is not influenced by the misalignment effects, under the
assumption that positive and negative muons have the same momentum spectrum.

Equation (\ref{eq:intbdl_approx}) can be used to relate the right side
of Eq.~(\ref{eq:delta}) with the average flux density in the yoke
plate. The systematic uncertainties deriving from the
assumptions used in writing these
two expressions can be suppressed normalizing Eq.~(\ref{eq:delta}) to
the bending expected from the field map:
\begin{equation}
\frac{[(\phi^{\rm prop}_{i+1}-\phi^{\rm data}_{i+1}) -
(\phi^{\rm prop}_{i}-\phi^{\rm data}_{i})]\cdot p_{T}}
{(\phi^{\rm prop}_{i+1}-\phi^{\rm prop}_{i})\cdot p_{T}} =
\frac{\langle B^{\rm map}_z\rangle-\langle B^{\rm true}_z\rangle}{\langle
  B^{\rm map}_z\rangle}|_{i+1/i}.
\label{eq:relDiscrepancy}
\end{equation}

Charge-antisymmetric combinations are computed
separately for the numerator and the denominator of the expression on
the left side. Each yoke plate is treated separately. The expression
on the right side represents the relative discrepancy of the
flux density averaged over a single yoke plate.

The ratio $\langle{B_z^{\rm true}}\rangle/\langle{B_z^{\rm map}}\rangle$ is computed using
Eq.~(\ref{eq:relDiscrepancy}).
It can be interpreted as the corrective scaling factor that has to
be applied to the $\vec{B}$ vector given by the magnetic
field map, in each point within the considered steel yoke plate, in order to
obtain the best estimate of ${B_z^{\rm true}}$ that reproduces
the measured track bending as observed in that plate.

\subsection{Results in the barrel yoke}
\label{sec:CRAFTBarrelResults}

This technique was applied on the sample of cosmic rays collected
during the CRAFT campaign~\cite{CRAFTGeneral}. Only runs with stable
magnet conditions at a central magnetic flux density of 3.8~T were
selected. The inner tracker and  the DT chambers were aligned using
cosmic muon tracks, survey measurements and optical
systems~\cite{CFTTkAlignment, CFTMuTrackAlignment, MuonHWAlign}.
DT segments reconstructed in the transverse plane
($R-\phi$) were required to include at least seven hits out of the
eight available measurement planes in a chamber.
Tracks to be used in the analysis were required to be reconstructed by
the inner tracker with at least 10 hits and to have a momentum within
$15 < p_T < 100$~GeV/$c$.
In addition, the extrapolated track path was required to pass through
all four DT stations in the same sector and wheel.
About one million tracks survive these pre-selection requirements.
%Actual statistics:
%                mu+        mu-         total
%tight cuts    336886     265923        602809
%preselection  585370     463871       1049241

For the computation of scaling factors to be used to calibrate the field
map, additional selection criteria were applied in addition to the
pre-selection requirements, to select data of the best quality. Only runs of
certified good quality by the Data Quality Monitoring system
for both the Drift Tube system and the inner tracker were used, and
tracks were selected with a transverse impact parameter with respect
to the centre of CMS, $|d_0|$, less than 0.4~m
and a longitudinal impact parameter, $|d_z|$, less than 1~m. This
tighter selection retains about 0.6 million events.
The reduction in the size of the event sample is mostly
due to the increased pointing requirement and, therefore, is
especially severe (about a factor 5) in wheels $\pm2$.
Scaling factors were computed using the field map described in
Section~\ref{sec:implementation}.
Results for all sectors of a given layer and wheel, as well as
results for opposite wheels, were found to be
compatible within statistical uncertainties, as expected given that the main
$\phi$- and $z$-asymmetric features are described in this map; they
have, therefore, been averaged.
The resulting factors, listed in Table~\ref{tab:scalingFactors}, are
adopted as correction factors for the magnetic flux density in the
map used in the CMS software for reconstruction and
simulation.
\begin{table}
\begin{center}
\caption{Scaling factors for the field map described in Section~\ref{sec:implementation}, averaged between
opposite wheels. Reported errors represent the statistical uncertainty only.
\label{tab:scalingFactors}}
\vspace{3mm}
\begin{tabular}{l|ccc}
\hline
 & wheels $\pm$2 & wheels $\pm$1 & wheel 0 \\
\hline
L1 & 0.99 $\pm$ 0.04 & 1.004 $\pm$ 0.004 & 1.005 $\pm$ 0.005 \\
L2 & 0.96 $\pm$ 0.02 & 0.958 $\pm$ 0.003 & 0.953 $\pm$ 0.003 \\
L3 & 0.92 $\pm$ 0.08 & 0.924 $\pm$ 0.003 & 0.906 $\pm$ 0.003 \\
\hline
\end{tabular}
\end{center}
\end{table}

The values measured with tracks show the same trend as observed
with Hall probe measurements in the barrel (first two columns of
Table~\ref{tab:HallSummary}), although a direct numerical comparison is not
possible, as the Hall probe measurements are performed in the gaps
between wheels
while the tracks probe the field inside the steel plates.

\begin{figure}
  \begin{center}
    \includegraphics[width=\textwidth]{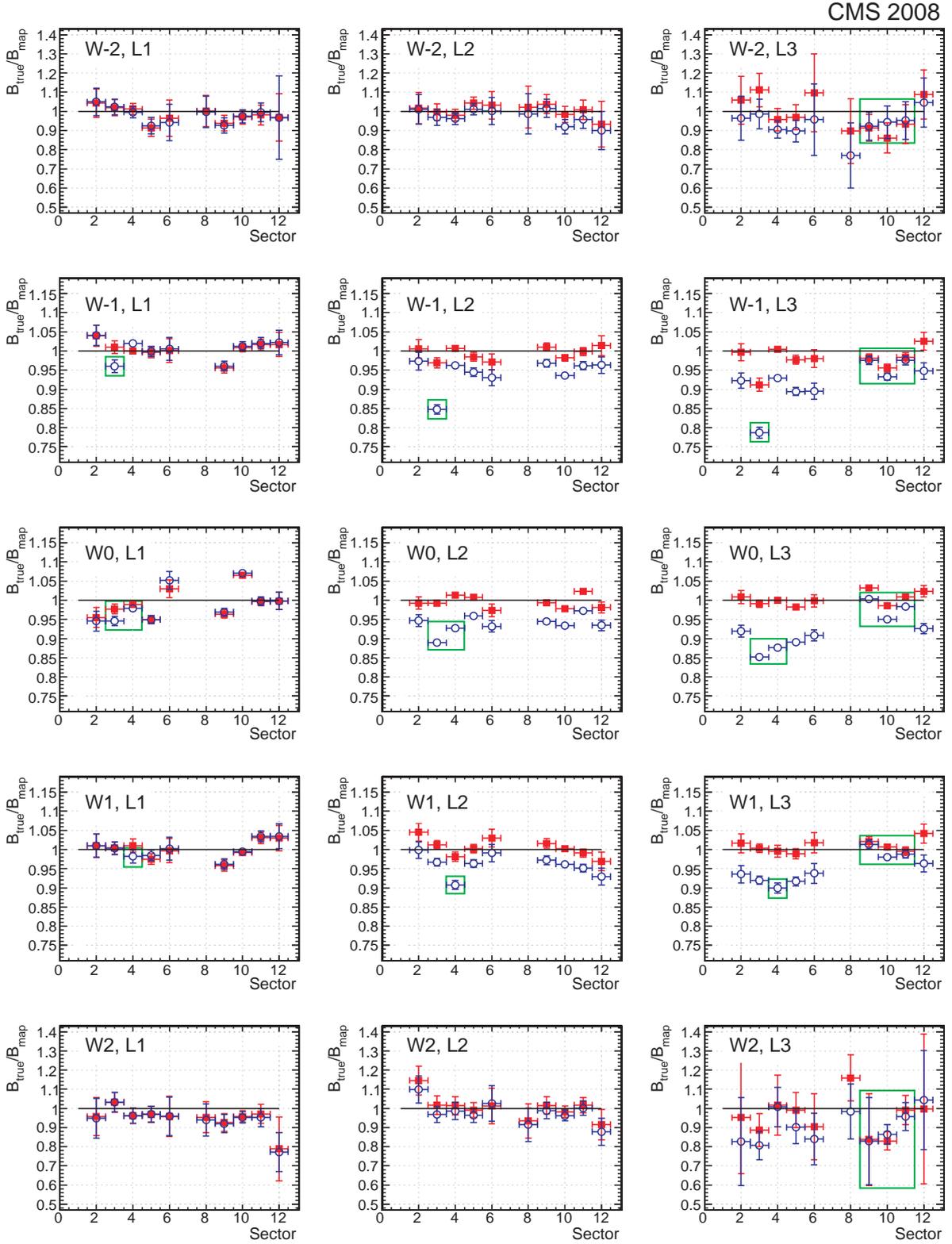}
    \caption{Solid squares: average values of
      $\langle{B_z^{\rm true}}\rangle/\langle{B_z^{\rm map}}\rangle$, after
      the correction factors of Table~\ref{tab:scalingFactors} are applied.
      Open circles: Scaling factors obtained using a
      12-fold $\phi$-symmetric map, with no correction and no special
      treatment of the $\phi$-asymmetric features described in
      Section~\ref{sec:implementation}. The sectors affected by these
      features have been highlighted with open boxes.
      Data points are not reported in sectors where the available
      event samples are too small
      (in particular, sectors around $y=0$, where the
      chambers are installed vertically). Error bars only include the
statistical uncertainty.
    \label{fig:scalingFactors}}
  \end{center}
\end{figure}

To verify the consistency of the method, the scaling factors were
recomputed using the corrected map, for events passing the
pre-selection criteria only.
The resulting factors, shown in Fig.~\ref{fig:scalingFactors}, are
compatible with unity within statistical uncertainties.
In order to search for possible biases they were averaged, grouping sectors in
different ways. The results are listed in
Table~\ref{tab:closureTest}. No bias is found to significantly exceed
the statistical uncertainties.
\begin{table}
\begin{center}
\caption{Consistency check: average values of $\langle{B_z^{\rm true}}\rangle/\langle{B_z^{\rm map}}\rangle$, after
  the correction factors of Table~\ref{tab:scalingFactors} are applied.
  Errors represent the statistical uncertainty of the method
  only.
}
\label{tab:closureTest}
\vspace{3mm}
\begin{tabular}{lc}
\hline
All sectors                  & 0.998 $\pm$ 0.001\\
\hline
All top sectors              & 0.997 $\pm$ 0.002\\
All bottom sectors           & 0.999 $\pm$ 0.002\\
%All sector 4                & 1.002 $\pm$ 0.002\\
%All sector 10               & 0.997 $\pm$ 0.002\\
All $x>0$ sectors            & 1.002 $\pm$ 0.003\\
All $x<0$ sectors            & 0.995 $\pm$ 0.003\\
\hline
All sectors (only L1)        & 0.997 $\pm$ 0.002\\
All sectors (only L2)        & 1.000 $\pm$ 0.002\\
All sectors (only L3)        & 0.997 $\pm$ 0.002\\
\hline
All sectors in wheel 0       & 1.000 $\pm$ 0.002\\
All sectors in wheels $\pm$1 & 0.998 $\pm$ 0.001\\
All sectors in wheels $\pm$2 & 0.984 $\pm$ 0.007\\
\hline
\end{tabular}
\end{center}
\end{table}
In particular, upper
and lower sectors agree well within statistical uncertainties indicating
that potential systematic biases due to the material budget description are
under control at the sub-percent level. A discrepancy in
the energy loss estimation should be visible as an
opposite bias in the top and bottom scaling factors,
given the propagation direction of the cosmic muons.
The 2.3\ $\sigma$ difference in the average for all sectors in wheels
$\pm$2 is due to the fact that the correction factors of
Table~\ref{tab:scalingFactors} were computed with a significantly
smaller sample because of the tighter selection, an effect particularly
important in these wheels.
%Therefore, the scaling factors were computed with a
%significantly smaller sample than the sample used in Table~\ref{tab:closureTest}.

To highlight the effect of the handling of the specific $\phi$-asymmetric
yoke features described in Section~\ref{sec:implementation},
the analysis was repeated on a $\phi$-symmetric map with no special
sector handling and no correction factors.
The results, for all events passing the pre-selection criteria, are shown
in Fig.~\ref{fig:scalingFactors}, separately for each
sector and yoke layer.
On top of the already observed overall scale bias in the
different layers, the effect of the chimneys is visible in sector 3 in
wheel $-1$, sector 3--4 in wheel 0, and  sector 4 in wheel $+1$. Also the
effect of the feet is visible in sectors 9, 10 and 11 of layer 3, where
scaling factors are higher than in the neighboring sectors. In these
locations, the true field integral is expected to be higher because of the
extra steel due to the feet supporting the yoke
(see Fig.~\ref{fig:Sectors}).
The handling of these features in the default map reduces their effect.

The analysis was cross-checked against a simulated
cosmic muon sample. Cosmic muons were
simulated with a Monte Carlo generator
(CMSCGEN~\cite{Biallass:2009ev}), interfaced
to the full GEANT simulation of the detector.
Only cosmic muons with a momentum of at least
10 GeV$/c$ at the entrance point in CMS were simulated.
%The cosmic muons were
%simulated with a flat arrival time distribution
%within $\pm$~12.5~ns to mymic the bunch crossing
%window of the trigger.
A realistic misalignment of the silicon tracker
and of the muon chambers corresponding to the CRAFT data was
applied to the reconstructed tracks and muon segments,
and the same pre-selection as used for the data was applied.
Since in the simulation ${B_z^{\rm true}}$ is identical
to ${B_z^{\rm map}}$ in every point in CMS by definition, this
study provides a powerful check of the calibration of the method.
The measured scaling factors have been verified to be compatible
with unity in all sectors.
%In Table~\ref{tab:scalingFactors_MC}, the
%resulting scaling factors are listed, averaged over sectors and wheels
%as was done in Table~\ref{tab:closureTest}.
The overall result shows that in the simulation,
without further calibration,
the method provides a scaling factor accurate to
better than a percent ($-0.7 \pm 0.4$\%). No evidence
is seen of any dependence versus radius (L1 -- L3),
versus $z$ (wheel~0 -- wheel~2), or propagation direction
(bottom versus top).

%\begin{table}[htbH]
%\begin{center}
%\caption{Scaling factors computed on Monte Carlo sample.
%\label{tab:scalingFactors_MC}}
%\vspace{3mm}
%\begin{tabular}{lc}
%\hline
%All sectors and wheels       & 0.993 $\pm$ 0.004\\
%\hline
%All top sectors              & 0.995 $\pm$ 0.005\\
%All bottom sectors           & 0.994 $\pm$ 0.006\\
%All sector 4                & 0.996 $\pm$ 0.007\\
%All sector 10               & 0.991 $\pm$ 0.007\\
%\hline
%All sectors (only L1)        & 0.998 $\pm$ 0.007\\
%All sectors (only L2)        & 0.992 $\pm$ 0.005\\
%All sectors (only L3)        & 0.991 $\pm$ 0.006\\
%\hline
%All sectors in wheel 0       & 0.993 $\pm$ 0.005\\
%All sectors in wheels $\pm$1 & 0.994 $\pm$ 0.005\\
%All sectors in wheels $\pm$2 & 0.98 $\pm$ 0.02\\
%\hline
%\end{tabular}
%\end{center}
%\end{table}

%%%%%%%%%%%%%%%%%%%%%%%%%%%%%%%%%%%%%%%%%%%%%%%%%%%%%%%%%%%%%%%%%%%%%%
\subsection{Systematic Uncertainties}
\label{sec:systematics}

As shown in the previous section, the available data sample of cosmic
muons can be used to constrain the scale of the magnetic field with a
statistical accuracy better than a percent in most of the barrel return yoke.
Possible systematic uncertainties in the method are
discussed in this section.
While the magnetic field map only provides a value of the $B$ vector in
every point, without an associated uncertainty, this information can be
used in physics measurements to derive systematic uncertainties due to
the
mismodeling of the magnetic field, by studying the effect of an
appropriately distorted field map.

A first systematic effect can arise from the assumption that
a good representation of the true magnetic field can be
obtained by applying the ratio
$\langle{B_z^{\rm true}}\rangle/\langle{B_z^{\rm map}}\rangle$ measured from
Eq.~(\ref{eq:relDiscrepancy}) as a
corrective scaling factor to the $\vec{B}$ vector in the
field map, in each point within the corresponding steel yoke plate.
A single scaling factor per steel block can correct for the average magnetic
field discrepancy, but not for its local variations within the block.

To quantify the possible magnitude of these variations, it is
useful to have a model of (realistic) distortions, somewhat larger
than the observed ones.
The TOSCA model computed using the smallest enclosing model boundaries of
those described in Fig.~\ref{fig:flux} ($R<13$~m and $z<20$~m) is
used for this purpose.
Reducing the volume boundaries forces more magnetic flux to return through
the yoke, and therefore provides a ``physical'' distortion of the
magnetic field map, still satisfying Maxwell's equations while
providing natural variations in all $B$ components
(unlike a single scaling factor for all field components in a given
region).
The magnitude of the distortion in this model is arbitrary. Given that
discrepancies with respect to the default model in the barrel yoke
layers are at least twice as large as the actual measured discrepancies
reported in Table~\ref{tab:scalingFactors}, all systematic effects
that are estimated from the comparison of this model with the default
one are scaled down by a factor 2.

The ratio between the value of $B$ predicted by the standard and distorted
models is computed as a function of the position inside a plate.
Its variation within the plate, that is found as expected to be much
smaller than the variation of $B$ itself, represents the variation
that the scaling factor would have in the different regions of the plate.
Since the method described in Section~\ref{sec:CRAFTBarrel} averages the
scaling factor in the plate, and over a region
that covers a limited $z$ range due to track selection criteria,
a systematic uncertainty is estimated as half of the maximum
observed variation between any two points within the
plate. This correspond to $\pm$0.5\%  in all layers and wheels,
except for layer 1 of wheels $\pm$1 and $\pm$2, where the
variations are $\pm$2.5\% and $\pm$2\%, respectively. This is the region
of the yoke where the field is the highest and closest to saturation,
giving rise to deviations from linear scaling.
This systematic uncertainty could be reduced by deriving
scaling factors for smaller regions in $z$ and $\phi$, at the cost of larger
statistical uncertainties.

Other systematic effects can arise from the assumptions, discussed in
Section~\ref{sec:CRAFTBarrel}, used to derive the
average scaling factors:

\begin{itemize}
\item Effect of $B_r$. The radial component of the field is neglected
  in the analysis, although it can affect the measured bending. This
  component is negligible in most of the barrel yoke, except in the
  two inner steel layers of wheel $\pm$2, where it reaches 0.4~T.
  %... can mention that this prevents to use Eq. 2 to get Bz.
  The presence of a $B_r$ component affects both the
  real muon bending and the track extrapolation, so to first
  order its effect cancels in their comparison and in the ratio of
  Eq.~(\ref{eq:relDiscrepancy}).
  A bias on the measured scaling
  factor is expected only if the ratio
  $B_r^{\rm true}/B_r^{\rm map}$ differs from the ratio
  $B_z^{\rm true}/B_z^{\rm map}$. The distorted field map described above was
  used as a model of a physical distortion of
  both components. The resulting bias on the measured scaling factor
  is estimated to be 0.5\% in L1 and L2 of W$\pm2$, and negligible elsewhere.

\item Assumption of $\phi$ and $z$-symmetry in averaging scaling
  factors.
  As described in Section~\ref{sec:implementation}, the current field map is
  implemented assuming 12-fold $\phi$-symmetry of the yoke, with
  a specific handling of sectors affected by the presence of
  chimneys, feet and carts. As a result of the specific description of
  these special sectors, the measured scaling factors
  for all sectors in each layer and wheel, as well as
  in opposite wheels, are compatible and can be averaged.
  However, residual differences can be present also among the sectors
  where symmetry is assumed to hold. Differences of up to
  $\pm$1\% in the field integral for a path traversing radially these yoke
  layers are predicted by the TOSCA model.
  A systematic uncertainty of $\pm$1\% due to the
  assumption of symmetries in the averaging of the scaling factors is
  therefore estimated.
\end{itemize}

Finally, the main input to the analysis is the muon segment angle measured
in each DT station, which can be affected by the imperfect knowledge of the
internal geometry of the chambers. Each chamber is composed of three groups
of layers of drift cells, called superlayers, that are superimposed~\cite{CMS}.
The assembly procedure can give an uncertainty  in the distances between
the superlayers of about 1~mm~\cite{PTDR1}
for a typical distance of approximately 25~cm.
The corrective scaling factors, $B_z^{\rm true}/B_z^{\rm map}$, are
evaluated from the difference between the angles measured in
consecutive stations and are, hence, particularly sensitive to a
correct description of their different internal geometries.
An uncertainty of 2\% on the scaling factors has been computed
assuming an unaccounted difference of 1~mm in the internal distance
between the two superlayers composing the chambers of two consecutive
stations.
This effect  is expected to vary from sector to sector as it is the result of
the interplay between the distribution of incident angles in a
given sector and the individual chamber geometry.

The statistical and systematic uncertainties on the measured scaling
factors are summarized in Table~\ref{tab:systematics}.

\begin{table}[htbH]
\begin{center}
\caption{Summary of statistical and systematic uncertainties on the
  measured scaling factors.
  The columns report the statistical uncertainty of the measurement and the
  estimated systematic uncertainties due to: the variation of the
  actual discrepancies within each plate; the neglected radial
  component of the magnetic field; the assumption of $\phi$ and $z$-symmetry
  in averaging scaling factors; and the internal geometry of the DT
  chambers. The last column gives the total uncertainty (sum in
  quadrature of the partial terms).
\label{tab:systematics}}
\vspace{3mm}
\begin{tabular}{c|cccccc}
\hline
Uncertainty (\%) & Statistical & Local variation & $B_r$ & Symmetry & Geometry & Total\\
\hline
W0, L1  & 0.5 & 0.5 & - & 1.0 & 2.0 & 2.3  \\
W0, L2  & 0.3 & 0.5 & - & 1.0 & 2.0 & 2.3  \\
W0, L3  & 0.3 & 0.5 & - & 1.0 & 2.0 & 2.3  \\
W$\pm$1, L1  & 0.4 & 2.5 & - & 1.0 & 2.0 & 3.8  \\
W$\pm$1, L2  & 0.3 & 0.5 & - & 1.0 & 2.0 & 2.3  \\
W$\pm$1, L3  & 0.3 & 0.5 & - & 1.0 & 2.0 & 2.3  \\
W$\pm$2, L1  & 3.8 & 2.0 & 0.5 & 1.0 & 2.0 & 4.8 \\
W$\pm$2, L2  & 2.4 & 0.5 & 0.5 & 1.0 & 2.0 & 3.4 \\
W$\pm$2, L3  & 7.7 & 0.5 & - & 1.0 & 2.0 & 8.0  \\
\hline
\end{tabular}
\end{center}
\end{table}

The scaling factors measured in the present study
are insensitive to the radial distribution of the field map discrepancy
inside steel blocks; they simply correct the integral of $B$ along the path.
For example, the two outermost barrel steel layers consist of a
sandwich of three steel plates with different magnetic properties; the
actual discrepancy of the map may be different in each one.
While such radial variations of the discrepancy with respect to the average
can affect the predicted position of a muon after traversing the plate,
they have no effect on the bending angle.

The measured average discrepancies do not
exclude larger localized discrepancies. Local deviations are, however,
constrained by the continuity of the magnetic flux.

\section{Conclusions}
\label{sec:conclusions}
The magnetic flux density in the steel plates of the CMS barrel return yoke was measured
precisely using cosmic ray muons, leading to a fundamental improvement
in the understanding of the CMS magnetic field.
The results are consistent with the indication of
measurements with Hall probes installed in the gaps between wheels.
Based on these measurements, an
improved map of the CMS magnetic field for the central magnetic flux
density planned for the first years of physics operation ($B_0=3.8$~T)
has been provided for simulation,
High-Level Trigger, and track reconstruction.
In the CMS yoke, the new map is estimated to be accurate to better
than 3\% in the steel of the three central barrel wheels, and to
about 8\% in the steel of the two outermost barrel wheels, satisfying the accuracy
required for physics analysis and muon triggering in CMS.
%The measurement described in this paper can be repeated using
%collision data, to possibly obtain an even higher precision and, if
%needed, to obtain corrections for the field maps at different coil currents.

\section*{Acknowledgments}
We thank the technical and administrative staff at CERN and other CMS
Institutes, and acknowledge support from: FMSR (Austria); FNRS and FWO
(Belgium); CNPq, CAPES, FAPERJ, and FAPESP (Brazil); MES (Bulgaria);
CERN; CAS, MoST, and NSFC (China); COLCIENCIAS (Colombia); MSES
(Croatia); RPF (Cyprus); Academy of Sciences and NICPB (Estonia);
Academy of Finland, ME, and HIP (Finland); CEA and CNRS/IN2P3
(France); BMBF, DFG, and HGF (Germany); GSRT (Greece); OTKA and NKTH
(Hungary); DAE and DST (India); IPM (Iran); SFI (Ireland); INFN
(Italy); NRF (Korea); LAS (Lithuania); CINVESTAV, CONACYT, SEP, and
UASLP-FAI (Mexico); PAEC (Pakistan); SCSR (Poland); FCT (Portugal);
JINR (Armenia, Belarus, Georgia, Ukraine, Uzbekistan); MST and MAE
(Russia); MSTDS (Serbia); MICINN and CPAN (Spain); Swiss Funding
Agencies (Switzerland); NSC (Taipei); TUBITAK and TAEK (Turkey); STFC
(United Kingdom); DOE and NSF (USA). Individuals have received support
from the Marie-Curie IEF program (European Union); the Leventis
Foundation; the A. P. Sloan Foundation; and the Alexander von Humboldt
Foundation.

\bibliography{auto_generated}

%%% Local Variables:
%%% mode: latex
%%% TeX-master: t
%%% End:
\cleardoublepage\appendix\section{The CMS Collaboration \label{app:collab}}\begin{sloppypar}\hyphenpenalty=500\textbf{Yerevan Physics Institute,  Yerevan,  Armenia}\\*[0pt]
S.~Chatrchyan, V.~Khachatryan, A.M.~Sirunyan
\vskip\cmsinstskip
\textbf{Institut f\"{u}r Hochenergiephysik der OeAW,  Wien,  Austria}\\*[0pt]
W.~Adam, B.~Arnold, H.~Bergauer, T.~Bergauer, M.~Dragicevic, M.~Eichberger, J.~Er\"{o}
% ...
\vskip\cmsinstskip
% ...

\end{sloppypar}
\end{document}